\documentclass[11pt,a4paper]{article}

\usepackage{fancyhdr,epsfig,url,lastpage}
\usepackage[dvips]{color}

\usepackage{amssymb,amsmath}
\usepackage{color}
\usepackage[english]{babel}
\usepackage{algorithm}
\usepackage{algpseudocode}
\usepackage[sort&compress]{natbib}
 \usepackage{graphicx}

\newcommand{\be}{\begin{equation}}
\newcommand{\ee}{\end{equation}}

\begin{document}

\thispagestyle{fancy}

\begin{center}

   \textbf{\Large Self Organization applied to Dynamic Network Layout}\\[5mm]
   \textbf{\large Markus M. Geipel}

   \begin{quote}
     \begin{center}
       Chair of Systems Design, ETH Zurich, Kreuzplatz 5, 8032 Zurich, Switzerland \\
       \texttt{mgeipel@ethz.ch}
     \end{center}
   \end{quote}

\end{center}
\begin{abstract}
  As networks and their structure have become a major field of research, a
  strong demand for network visualization has emerged.  We address this
  challenge by formalizing the well established spring layout in terms of
  dynamic equations. We thus open up the design space for new
  algorithms. Drawing from the knowledge of systems design, we derive a
  layout algorithm that remedies several drawbacks of the original spring
  layout.  This new algorithm relies on the balancing of two antagonistic
  forces. We thus call it {\em arf} for ``attractive and repulsive
  forces''.  It is, as we claim, particularly suited for a dynamic layout
  of smaller networks ($n < 10^3$). We back this claim with several
  application examples from on going complex systems research.

\end{abstract}

\section{Introduction}
\label{sec:intro}

In many disciplines - sociology, physics, mathematics, economics, biology
- networks and their structure have become a major concern
\cite{barabasi03, strogatz01, kauffman95}. So far, research focused
mostly on statistical properties of these networks. A new challenge is
the analysis of dynamic aspects of networks. In such networks, links are
created and vanish, nodes are added and dropped.  For both, the analysis
of the structure and the dynamics of networks, visualization is an
invaluable tool.

A very appealing and successful approach to the network layout problem is
the so called {\em force directed layout}. It was first proposed in 1984
\cite{eades84} and is related to the concept of self organization based
on only local interactions.  In fact, many natural structures become
functional, efficient and even visually appealing by self organization.
Often balancing two antagonistic forces gives rise to emergent order.
Balancing for example activation and inhibition in an artificial
chemistry was use to mimic the formation process of patterns on sea
shells \cite{meinhard03}.  Another example is the modeling of vortex
swarming of Daphnia \cite{mach06}: attractions ensure the coherence of
the swarm while repulsion prevents it from collapsing. Finally, a force model can be employed to position soccer
playing robots on the field: The robots feel attracted to the ball and
the goal and feel repelled by other robots \cite{veloso98}.  Similarly, a network can be regarded as a system comprised of agents
represented by the nodes and their interactions represented by links
between them.  In {\em force directed layout}, a combination of simple
forces leads to the emergence of global spatial structure.

The most commonly used class of such algorithms is the spring
layout.  It was for example used to dynamically explore the structure of
the WWW
\cite{brandes00}.  In the spring model, nodes experience a spring force
(Hooke's Law), that adjusts the distances of connected nodes and a
repulsion (Coulomb's Law) to spread out unconnected nodes. The mechanical
equilibrium is supposed to posses favorable properties like low edge
crossings and nearly equal edge length. It can be searched in two
different ways. The first option is to simply simulate the system. This
is also the only way to generate dynamic network layout. The second
option is to search for a local energy minimum more directly by general
global optimization methods. As we deal with dynamically changing
networks, we will only consider the first method.

\begin{figure}[htb]
  \centering
  \begin{tabular}{c c c}
    \includegraphics[width=0.3\textwidth]{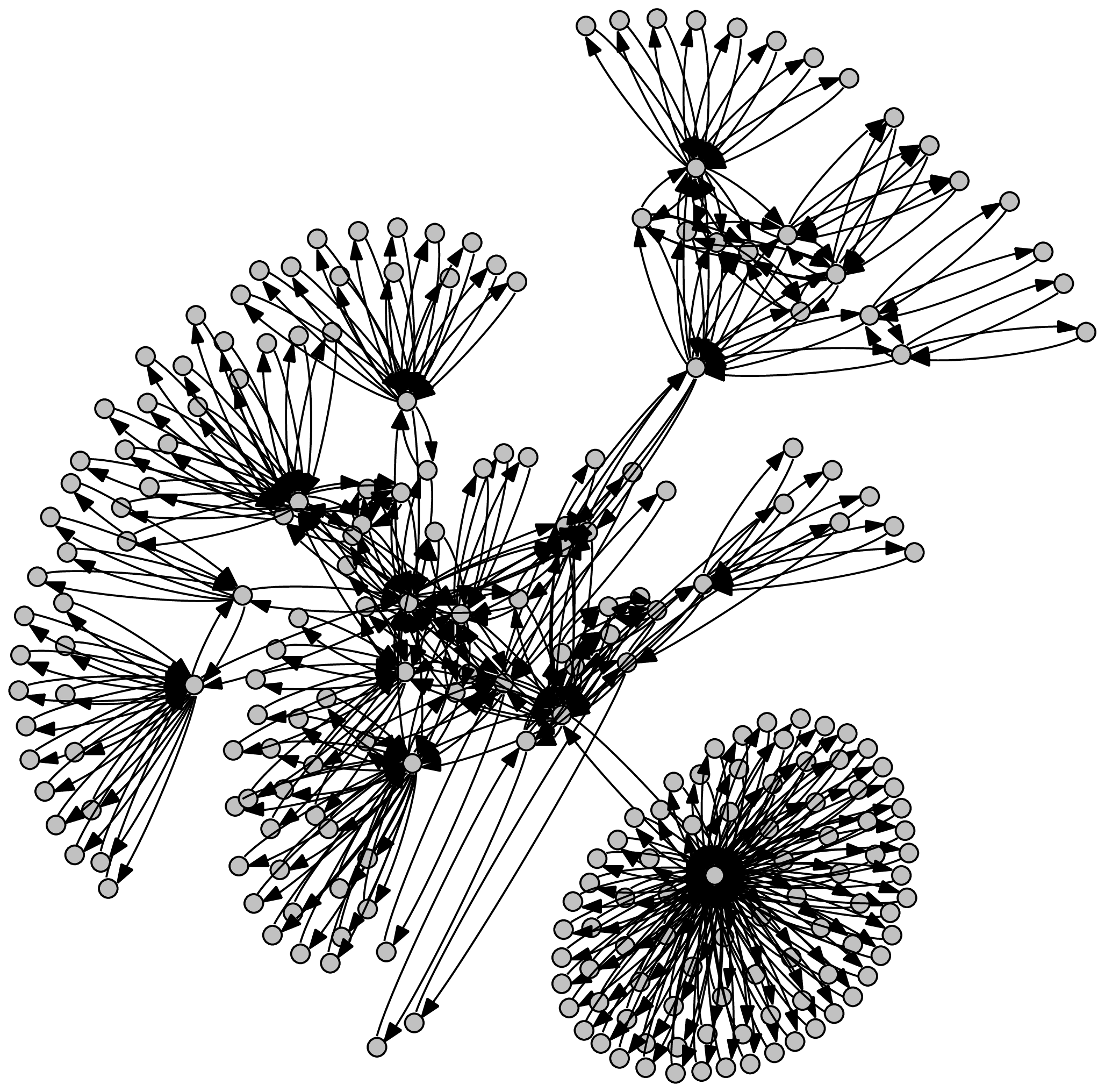}&
    \includegraphics[width=0.3\textwidth]{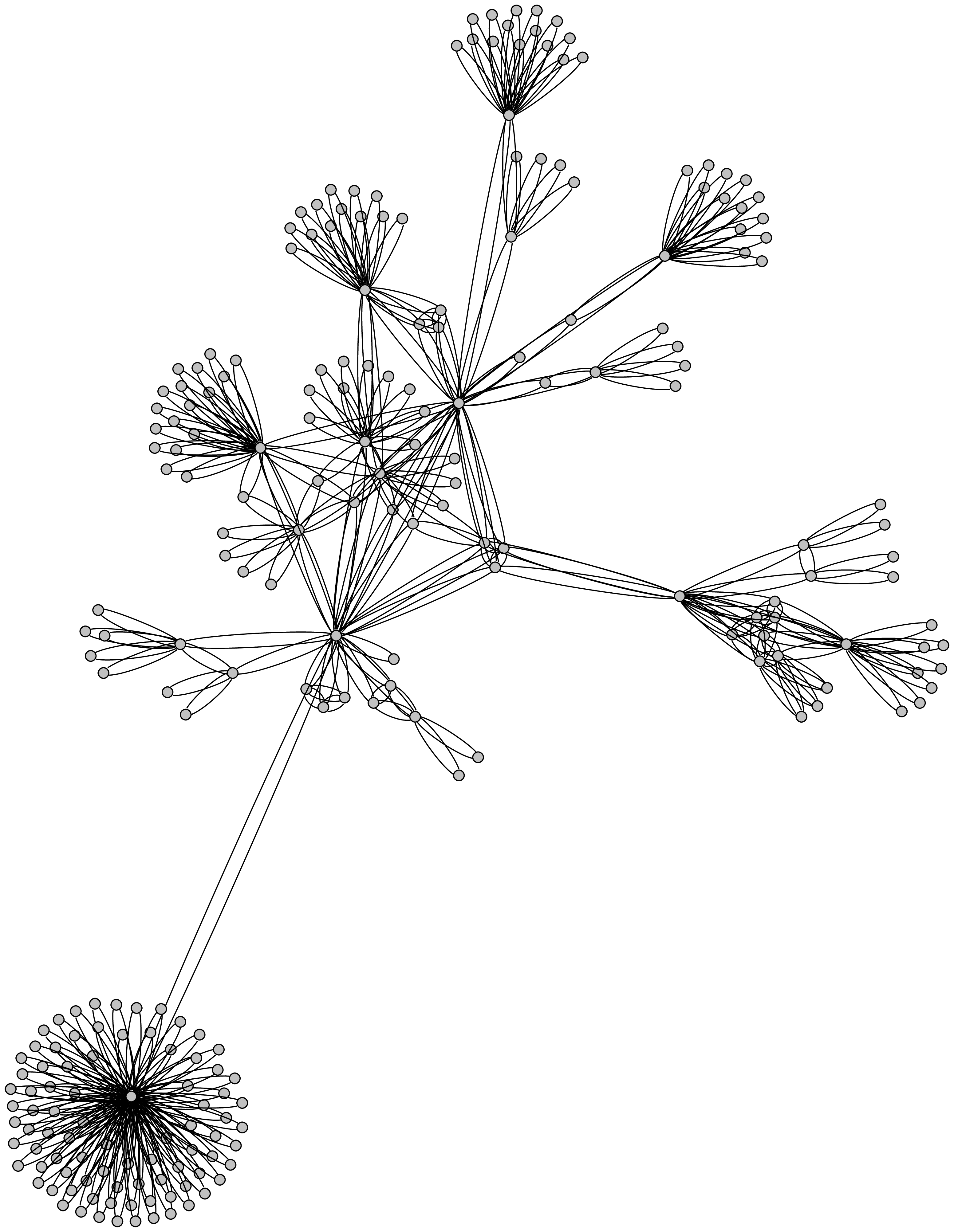}&
    \includegraphics[width=0.3\textwidth]{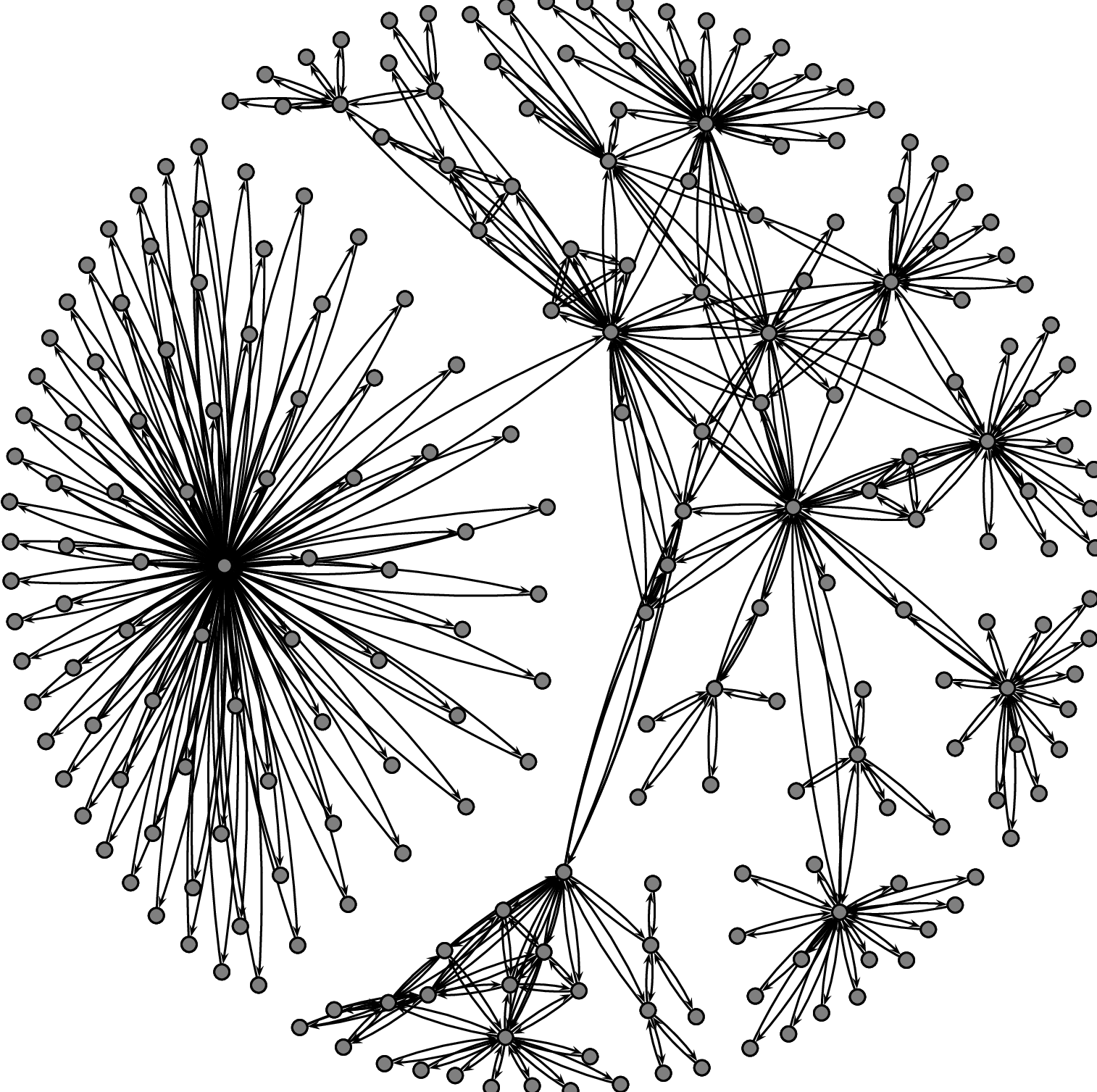}\\
    a & b &c\\
  \end{tabular}
  \caption{Network showing the relationships of managers based on their
    board memberships. Layout of the same network with {\em neato} on the
    left (a) and layout by {\em arf} on the right (c). For clarity
    reasons, labels where omitted.}
  \label{fig:manager}
\end{figure}

The most prominent representatives of spring layout are the Kamada-Kawai
\cite{kamada89} and the Fruchterman-Reingold \cite{fruchterman91}
algorithm. Figure \ref{fig:manager} a and b show a complex heterogeneous
network visualized with them.  Several problems spring to mind:
congestion, order less arrangement of nodes in the highly connected core
and high variation in the edge length.

The contribution of this paper is first, to analyze the shortcomings of
the spring model and second, to present a modified model that addresses
these shortcomings.  We will proceed as follows: First, we will formalize
and analyze the spring model in its most general from. Having opened up
the design space we propose changes to remedy the spring model's defects.
So, a new model will be derived from the general spring model.  In the
next section, an evaluation of the new model will be presented.  This
will be followed by several application examples in the field of complex
systems.  Finally concluding remarks will round up the discussion.

\section{The Spring Model: A Critical View}
\label{sec:springmodel}

For the following discussion we will assume a graph $G$ to be a tuple
$G=(V,E)$ where $V$ is the set of vertices or nodes respectively and $E
\subseteq V \times V$ is the set of edges. We will use the indices $i$
and $j$ to denote vertices. Whether the graph is directed or undirected
is irrelevant for the discussion.

The general spring model is based on the idea that springs adjust the
distances between connected nodes: connected nodes should be placed
equidistantly. The spring pushes towards this desired distance $d$ which
is the resting position of the spring.  The force on $i$ of a spring
between $i$ and $j$ is given as:
\begin{equation}
  \label{eq:spring}
  m \ddot{x}_i=   K_{i,j}{(\|x_j-x_i\|-d)}{(x_j-x_i)\over \|x_j-x_i\|} 
\end{equation}
$x_i$ is used to denote the position vector of node $i$.  $K_{i,j}$ is
the spring constant of the spring connecting $i$ and $j$.  Thus $K$ is
defined by:
\begin{equation}
  \label{eq:kspring}
  K_{i,j} = 
  \begin{cases}
    k, &\text{if $(i,j) \in E \lor (j,i) \in E$;}\\
    0, &\text{otherwise.}
  \end{cases}
\end{equation}

As only connected nodes have springs between them each node is endowed
with a repulsive field to adjust their
distances. The resulting
force experienced by $i$ in the presence of $j$ is thus:
\begin{equation}
  m \ddot{x}_i = - \rho{(x_j-x_i)\over \|x_j-x_i\|^{1+\beta}}
\end{equation}
Assuming, that we mimic a Coulomb field, $\rho$ can be interpreted as the
electrostatic constant multiplied by the charges of the two nodes.
$\beta$ is commonly set to $2$.

Unfortunately, a system like this will start oscillating. Therefore
friction is introduced:
\begin{equation}
  \label{eq:friction}
  m \ddot{x}_i= -\gamma \dot{x}
\end{equation}
$\gamma$ denotes the friction constant.

Putting the forces together we get the following equation:
\begin{equation}
  \label{eq:general2f}
  m \ddot{x}_i= -\gamma \dot{x} + \sum_{j \in
    V}K_{i,j}{(\|x_j-x_i\|-d)}{(x_j-x_i)\over \|x_j-x_i\|} - \rho \sum_{j \in V}{(x_j-x_i)\over \|x_j-x_i\|^{1+\beta}}
\end{equation}

under the assumption that $m \ll \gamma, \rho, K$ and $\gamma = 1$ we can
derive the following approximation for the movement of the nodes.

\begin{equation}
  \label{eq:general2v}
  \dot{x}_i  =  \sum_{j \in
    V}K_{i,j}{(\|x_j-x_i\|-d)}{(x_j-x_i)\over \|x_j-x_i\|} - \rho \sum_{j \in V}{(x_j-x_i)\over \|x_j-x_i\|^{1+\beta}}
\end{equation}

Having presented the general formulas for spring layout we will take a
look at the outcome.  Figure \ref{fig:manager}a shows a network,
visualized with the spring layout by Kamada and Kawai \cite{kamada89} as
implemented in graphvis' {\em neato} layouter. Figure \ref{fig:manager}b
shows the same network, visualized with an alternative spring model
proposed by Fruchterman and Raingold \cite{fruchterman91} (graphvis' {\em
  fdp} layouter).  We notice several problems: First, there is congestion
around strongly connected nodes (center area and bottom right).  To many
nodes are crammed in too little space by a force too strong. Unfortunately,
the strongly connected nodes are often the ones that are of special
interest in network analysis.  Second problem: the available layout space
is used inefficiently: there are big empty spaces, while in the center
some nodes are occluded by others in the case of figure
\ref{fig:manager}a. An even distribution of nodes would moreover
facilitate any
labeling of nodes. For figure \ref{fig:manager}b the structure is
clearer. However the star cluster is pushed too far off by its repulsive
field. The one connecting spring is not sufficient to compensate the
repulsion.  This unbalance is especially distinct in networks with highly
heterogeneously node degress such as the ones predominant in complex
systems research. Finally, the layout does not show much symmetry or structure.
Instead, especially in the center, leaves a rather chaotic impression. 
The self organizing forces do not seem strong
enough to disentangle the network. However, for laying out dynamically
changing graphs, this would be vital.

The basic problem is this: The different forces at work are not well
tuned. In some situations, attraction is to strong, in others repulsion
is too dominant.

\section{A New Model}
\label{sec:algorithm}

The new model centers around the principle of balancing two forces: An
attractive and a repulsive one. We thus call it {\em arf} We search for
forces, that better represent commonly accepted principles of graph
design like the following ones \cite{fruchterman91}:
\begin{enumerate}
\item Vertices connected by an edge should be drawn near each other.
\item Vertices should not be drawn {\em too} close to each other.
\end{enumerate}
The fact that``[...] the layout should display as much symmetry as possible''
 as noted by \cite{eades84} will be taken into account.  We will address these
issues by challenging assumptions made in the spring layout. In doing
so, we will derive a new force model for laying out networks.

Let us first take a look at the spring force: Obviously, connected nodes
should be close to each other for two reasons: First humans find it
difficult to follow long edges. Thus, short ones make the graph more
readable.  Second the shorter an edge the lower the risk of crossing
another one. Thus an attractive spring force between nodes makes perfect
sense. But, all nodes of a graph should stay close together to ensure
that they are evenly spread. This will avoid runaway clusters such as
the one in figure \ref{fig:manager} b because each repulsive connection
between two nodes is now {\em balanced} by an attractive one.  The new
equation for the spring constant $K$ is defined as follows:
\begin{equation}
  \label{eq:omega}
  K_{i,j} =
  \begin{cases}
    0, &\text{if $i = j$;}\\
    a, &\text{if $(i,j) \in E \lor (j,i) \in E$;}\\
    1, &\text{otherwise.}
  \end{cases}
\end{equation}
The parameter $a$ gives the strength of springs between connected nodes.
It has to be greater than one: $a>1$. The greater $a$, the clearer the
separation of unconnected sub clusters.

As described in the previous section, the resting position $d$ reflects
the desired edge length. The spring force can be both attractive and
repulsive, depending on the position of the nodes. {\em However,} there
is already a repulsive field: We have two possible repulsive
forces. By setting $d=0$ in equation (\ref{eq:spring}), we accomplish two things: First, we clearly
separate the attractive and repulsive forces. A more straight forward
force system is easier to tune.  Second, we abandon the rule to enforce
equal length edges which led to congestion around highly connected nodes.
The new equation for the attractive force is simply
\begin{equation}
  \label{eq:attraction}
  m\ddot{x}_i= \sum_{j \in (V \setminus i)} 
  {  K_{i,j}({x}_j - {x}_i)  }
\end{equation}.

Besides the concern to have connected nodes close together and edges to
be short, also an equal distribution of nodes in the available layout
space is desirable.  Repulsive movement addresses this concern. The
balance between attraction and repulsion regulates the edge length in a
very flexible way. However, in the previous section we diagnosed a miss
balance in the force system. Loosely connected clusters were driven away
by too much repulsion and in highly connected clusters, repulsion was not
strong enough to counter congestion. With the new function for $K$ we
already addressed this issue. Yet, we can still improve by changing
repulsion from a quadratically decaying force to a distance invariant
one. This is done by choosing $\beta=0$.  The following term is used to
calculate repulsion:
\begin{equation}
  \label{eq:repulsion}
  m\ddot{x}_{i}= 
   - \rho \sum_{j \in (V \setminus i)} { 
    { {x}_j - {x}_i \over { \|{x}_j -
        {x}_i} \|}}  
\end{equation}

Next, we have to take into account changing graph sizes. The node density
should be invariant, not the size of the layout space. To scale the
layout space based on the number of nodes in the graph, the repulsive
force is multiplied by $\sqrt{|V|}$.  This is accomplished by defining
\begin{equation}
  \label{eq:rho}
  \rho = b\sqrt{|V|}
\end{equation} in equation (\ref{eq:repulsion}).
$b$ in equation (\ref{eq:rho}) scales the radius
of the layout space. 

When combining the force equations (\ref{eq:attraction}),
(\ref{eq:repulsion},\ref{eq:rho}), and the friction we can derive the
following movement approximation for the nodes:
\begin{equation}
  \label{eq:all}
  \dot{x}_i = \sum_{j \in (V \setminus i)}(K_{i,j} - {b \sqrt{|V|}  \over
    { \|{x}_j - {x}_{i}\|}}) ({x}_j - {x}_{i})
\end{equation}
The equation always leads to a circular layout space for the graph which
is also very convenient for generating {\em animations} as frame sizes
are invariant.  Node density can be adjusted with the parameter $b$ and
the separation of unconnected nodes with the parameter $a$ in the
equation for $K$.  The complete layout procedure is given by algorithm
(1).
\begin{algorithm}[htb]
  \label{alg:layout}
  \caption{ARF Layout($G(V,E)$)}
  \begin{algorithmic}[1]
    \Repeat \State $error \gets 0$ \For {{\bf all} $i \in V$} \State
    $\dot{x}_i = \sum_{j \in (V \setminus i)}{ (K_{i,j}+ {b {\sqrt{|V|}}
        \over { \|{x}_j - {x}_{i}}\|}) ({x}_j - {x}_{i}) }$ \State ${x}_i
    \gets {x}_i + \Delta t {\dot{x}}_i$ \State $error \gets error +
    |\dot{x}_i|$
    \EndFor
    \Until $error < \epsilon$
  \end{algorithmic}
\end{algorithm}
The update of the node-position takes place in line 5 of algorithm (1).
The parameter $\Delta t$ adjusts the size of one time step and thus the
precision of the numerical integration.  The choice of  $\Delta t$  is a
tradeoff between  
speed and potential instability which can be addressed by making $\Delta
t$ depending on $|\dot{x}_i|$.  The whole
update is repeated until the amount of changes in the graph drops beneath
a certain threshold $\epsilon$.

\section{Evaluation}
\label{sec:evaluation}

In this section we provide an evaluation of the new layout rules.  As
there is no commonly accepted benchmark for network layout. The following
evaluation cannot and does not attempt to provide hard evidence. It
rather attempts to indicate advantages of the movement model in a clear
cut situation: layout of small to medium size networks.
  
As reference point for comparisons, the layouters contained in the {\em
  graphviz} package\footnote{available under the Common Public License at
  \url{http://www.graphviz.org/}} were used. {\em Graphviz} is maintained
by AT\&T and probably the best currently available graph layout tool.  It
contains the following five layouters: {\em dot} \cite{gansner93}, {\em
  neato} \cite{kamada89}, {\em circo} \cite{Six99,kaufmann02}, {\em
  twopi} \cite{wills97} and {\em fdp} \cite{fruchterman91}.  The
following evaluation compares {\em arf} in respect to layout quality.
Com\-putational complexity is not the issue here.  Suffice it to say that
all kinds of force directed layouts reside in the same complexity class, which is
assumed to be $O(|V|^2)$%
\footnote{There exist techniques to achieve even lower
complexity. For more detailed discussions see
  \cite{fruchterman91,kamada89,eades84,davidson96,eades00a}.}.

When assessing layout quality, the following four criteria are commonly
accepted:
\begin{enumerate}
\item minimal number of crossing edges
\item aesthetics of the layout
\item separation of clusters
\item usage of the available space
\end{enumerate}

\begin{figure}[h]
  \centering
  \begin{tabular}{|c|c|c|c|}
    A&B&C&D\\
    \hline
    \includegraphics[width=2.5cm]{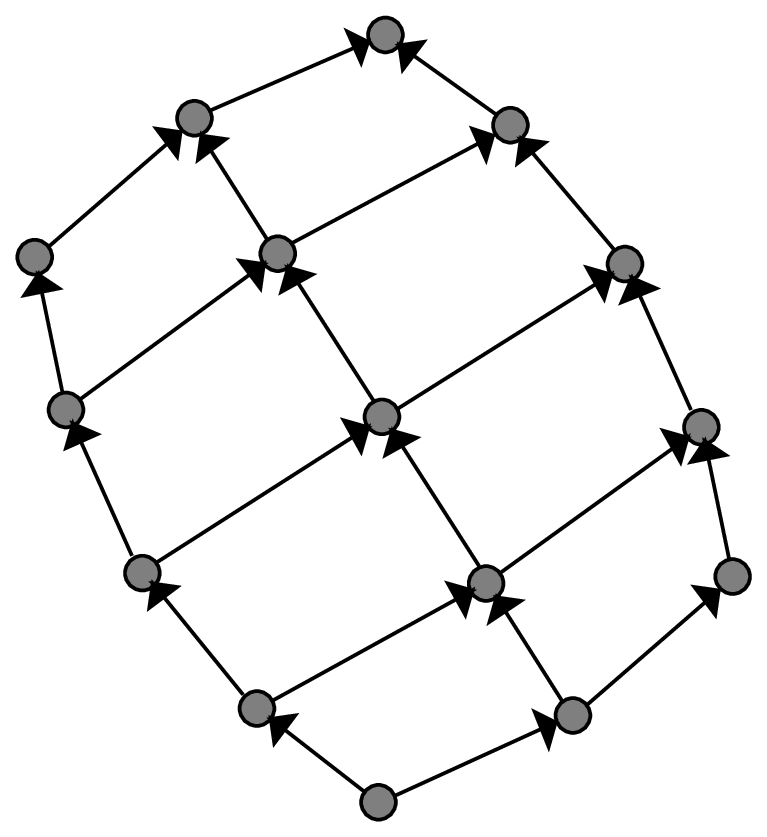} & \includegraphics[ width=2.5cm]{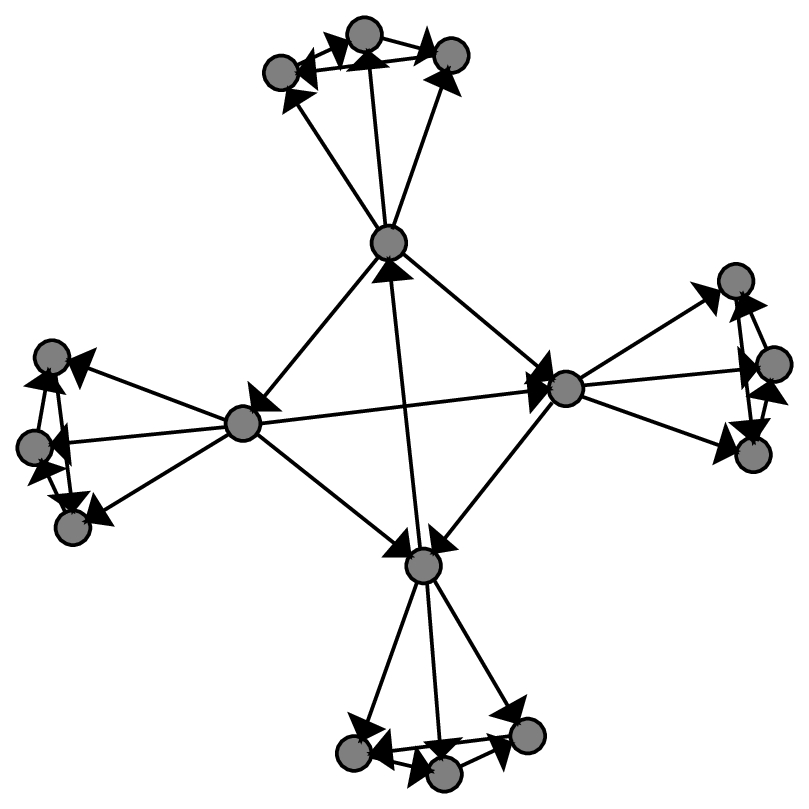}&
    \includegraphics[width=2.5cm]{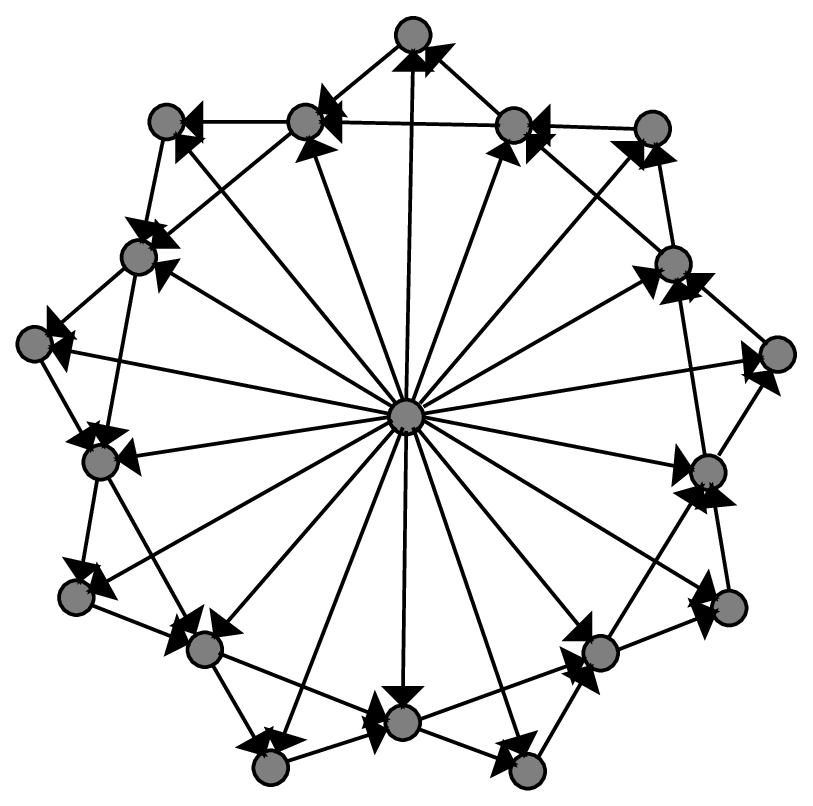} & \includegraphics[ width=2.5cm]{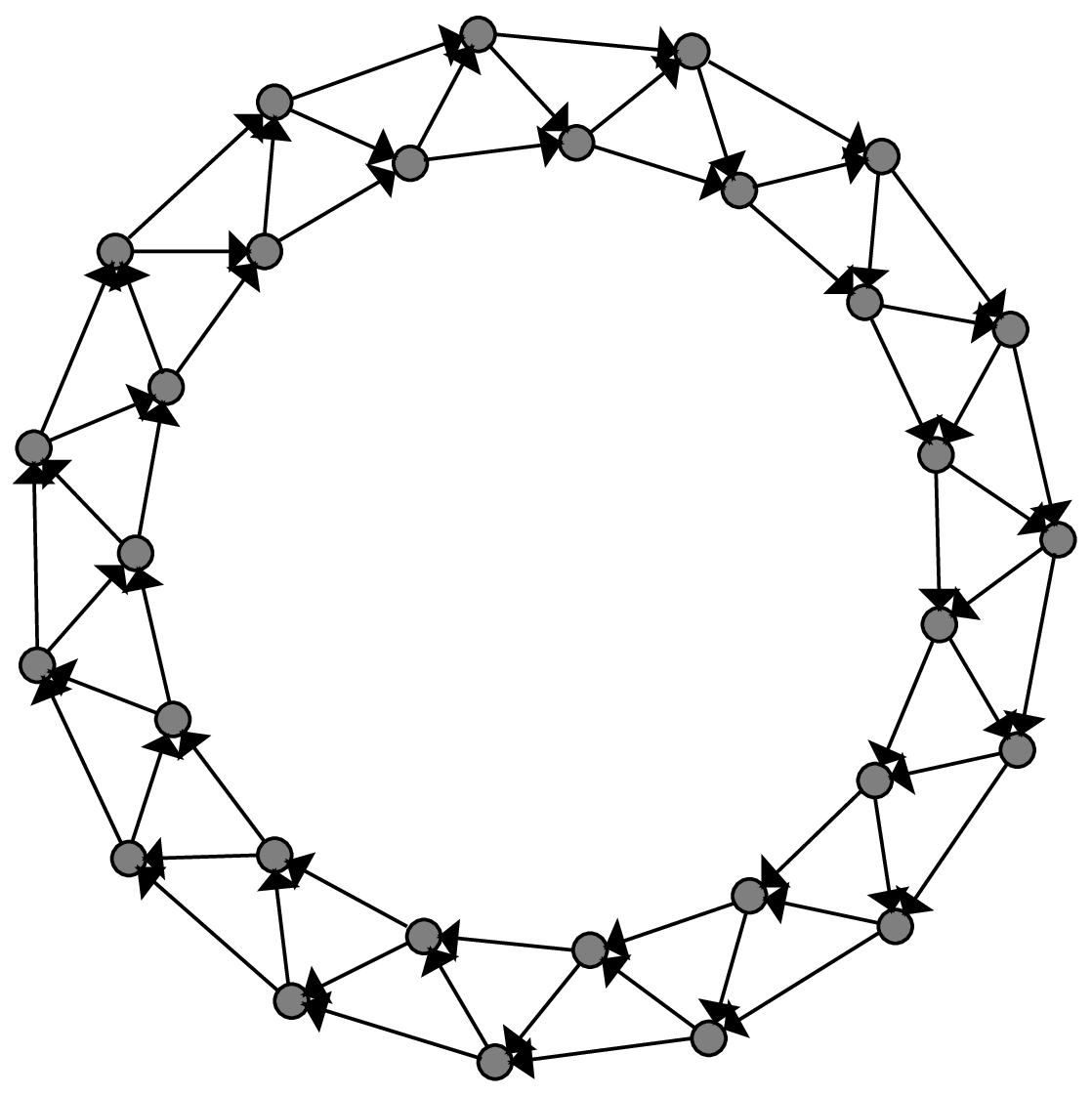} \\
    \hline
  \end{tabular}

  \begin{tabular}{ccc}
    \includegraphics[width=3.5cm]{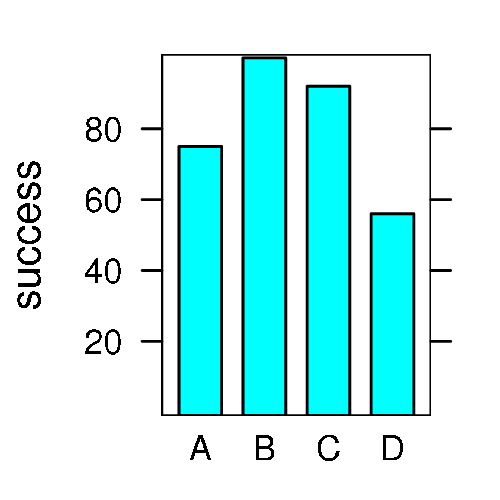} & 
    \includegraphics[width=3.5cm]{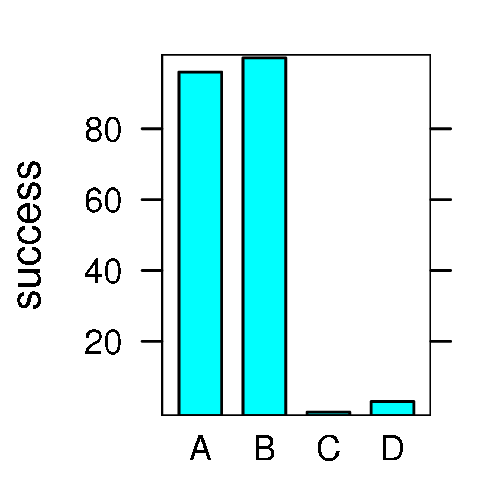}& 
    \includegraphics[width=3.5cm]{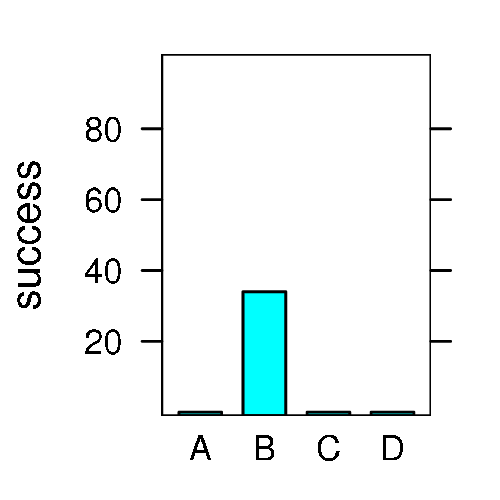}\\
    arf & neato & ftp \\
  \end{tabular}

  \caption{Top: the four graphs (A,B,C and D) where used as benchmark.
    Bottom: Percent of successful layouts of each layouter for each of
    the benchmark graphs (number in percent).}
  \label{fig:success}
\end{figure}

First, we will address points one and two: edge crossings and aesthetics.
As the initial node positions are random, just looking at {\em one}
generated layout is not enough. Thus, the following benchmark was used:
The task is to lay out the four symmetrical graphs shown in figure
\ref{fig:success} one hundred times.
The layouts in the figure were produced by {\em arf}.  Besides {\em arf},
the evaluation takes into account the layouters {\em neato} and {\em
  fdp}, both of which are probabilistic and force directed, too.  The
criteria for success are quite simple: For graphs A and D, each symetric
layout without crossings is counted as success.  For graphs B and C, no
algorithm finds a crossing-free layout. So, all symetric layouts, that
have no more crossings than the best produced layout, will be accepted.
Figure \ref{fig:success} shows the results.  It can be seen that {\em
  arf} applied to a random node distribution, in average, finds the
optimal layout more often than {\em neato} and {\em fdp}, even though
they use more sophisticated, non-incremental optimization algorithms.
The only exception is the grid graph (A) where {\em neato} outperforms
{\em arf}.  For graphs C and D however, {\em arf} performs better.
Please note that this benchmark only aims at small graphs. And the
selection of only four graphs in not representative.

\begin{figure}[tb]
  \centering
  \begin{tabular}{ c|c|c|c|c|}

    &E&F&G&H\\
    \hline {\bf arf} &
    \includegraphics[height=1.7cm, width=1.7cm]{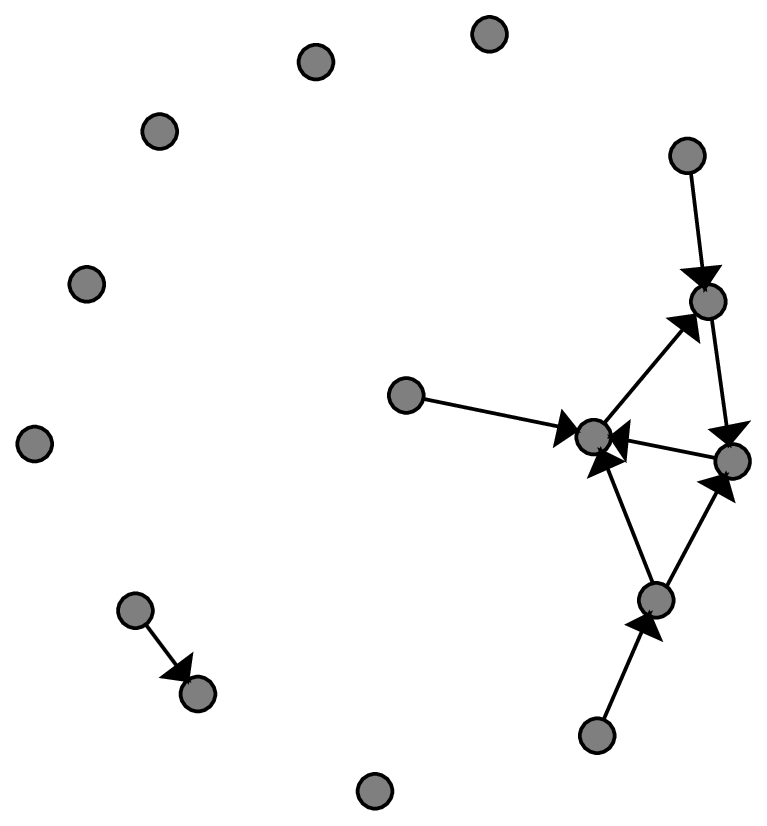} &
    \includegraphics[height=1.7cm, width=1.7cm]{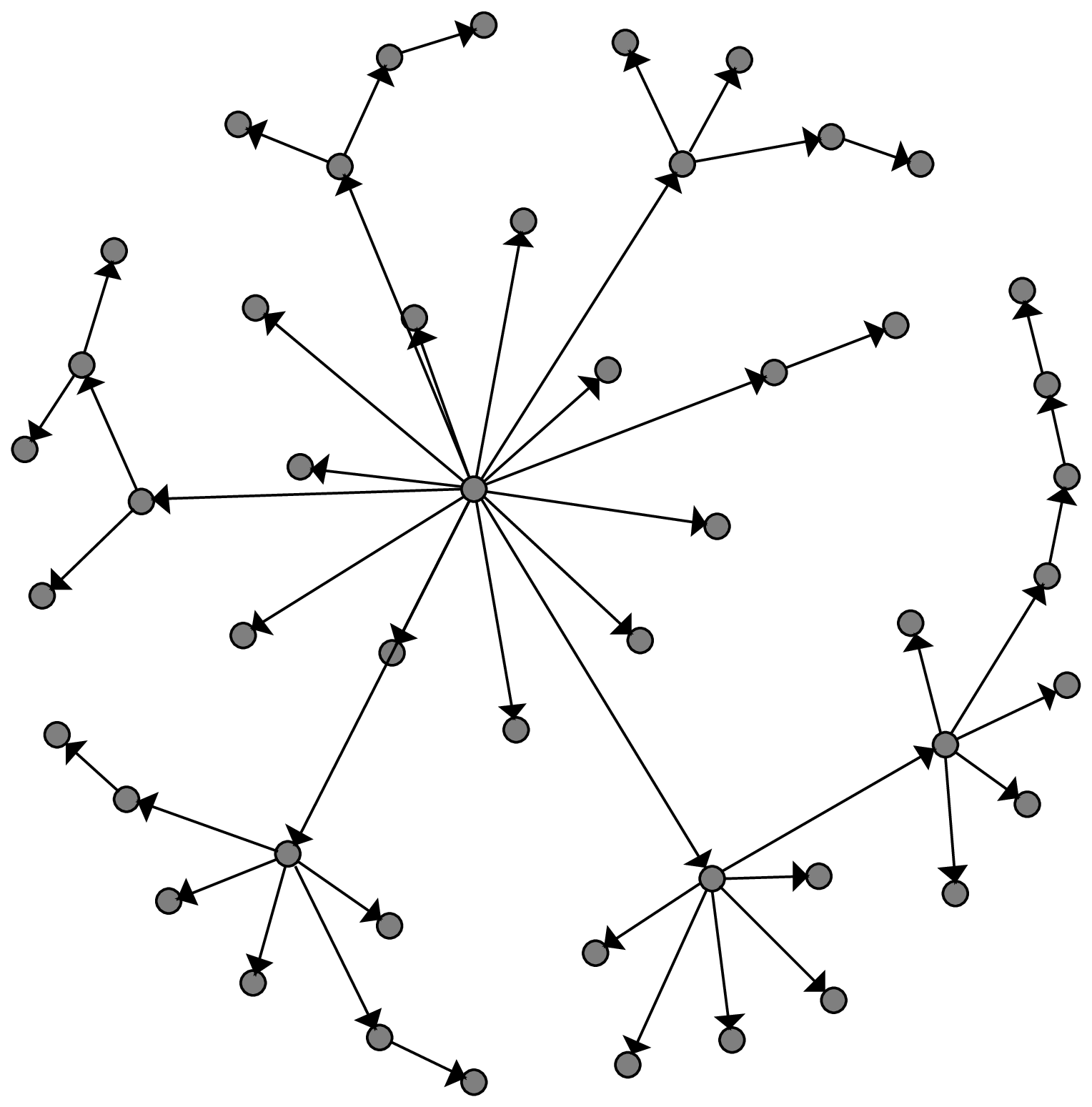} &
    \includegraphics[height=1.7cm, width=1.7cm]{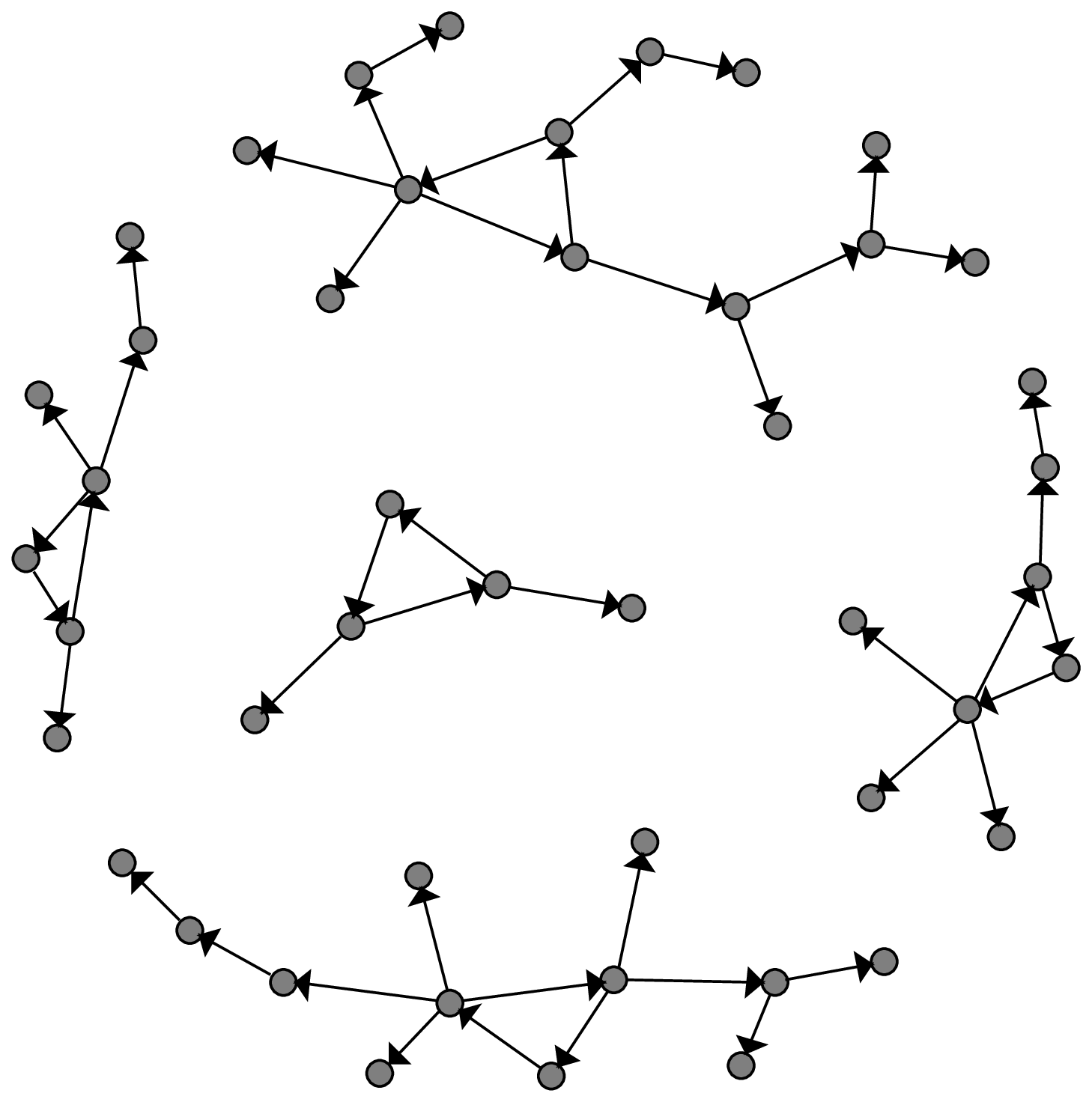} &
    \includegraphics[height=1.7cm, width=1.7cm]{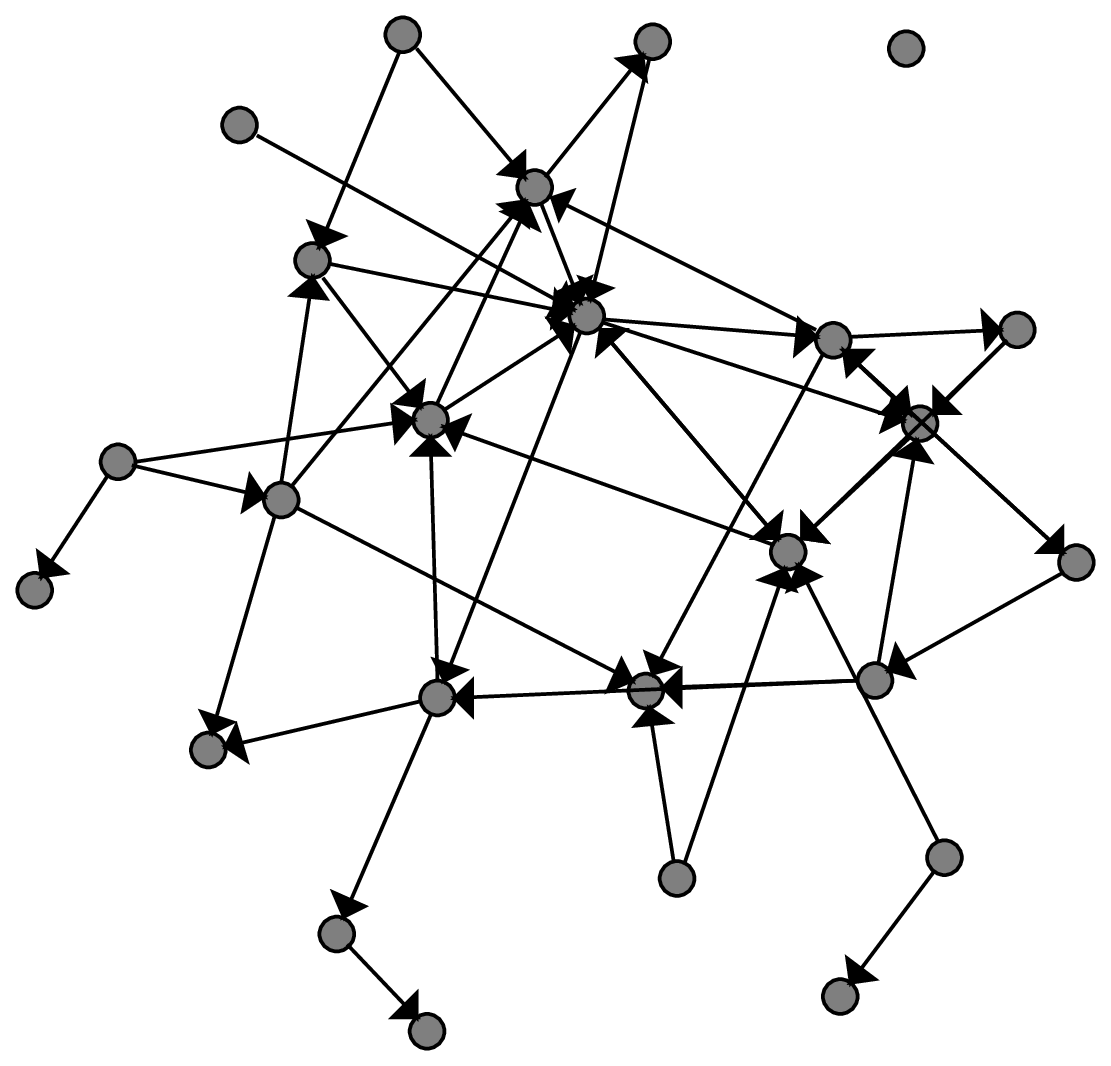} \\
    \hline
    { dot} &
    \includegraphics[height=1.7cm, width=1.7cm]{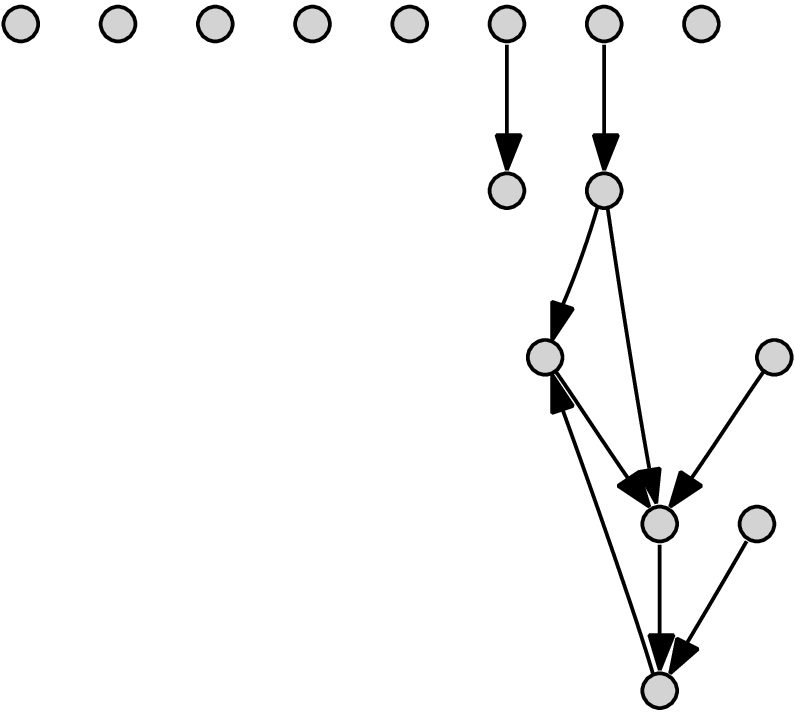} &
    \includegraphics[width=1.7cm]{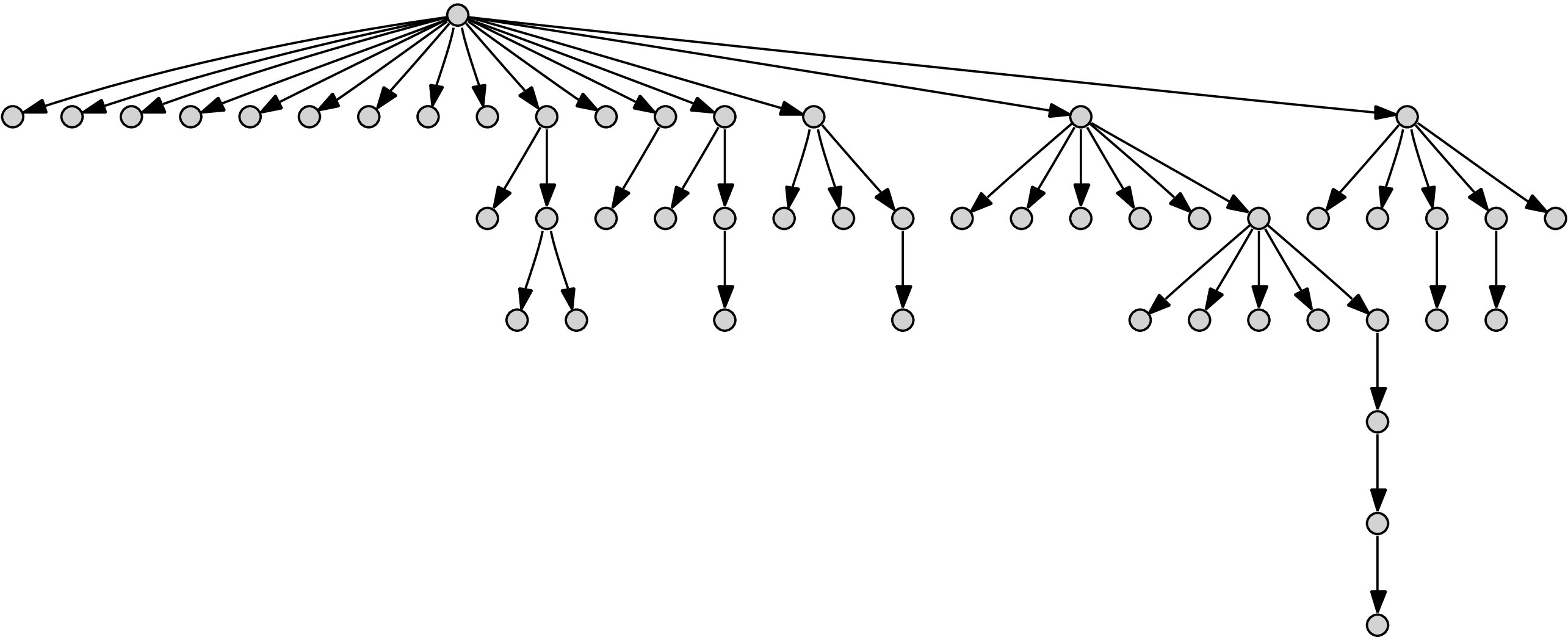} &
    \includegraphics[ width=1.7cm]{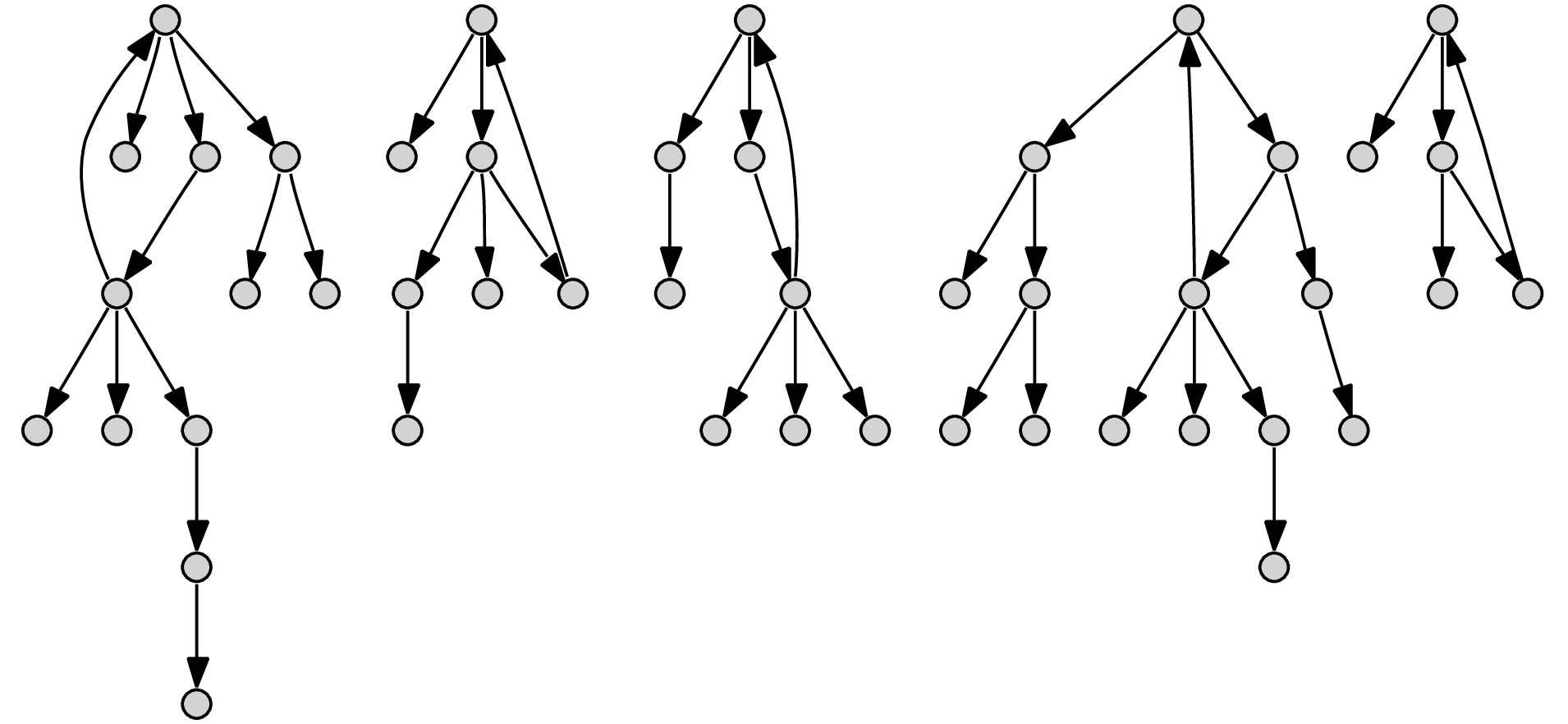} &
    \includegraphics[height=1.7cm]{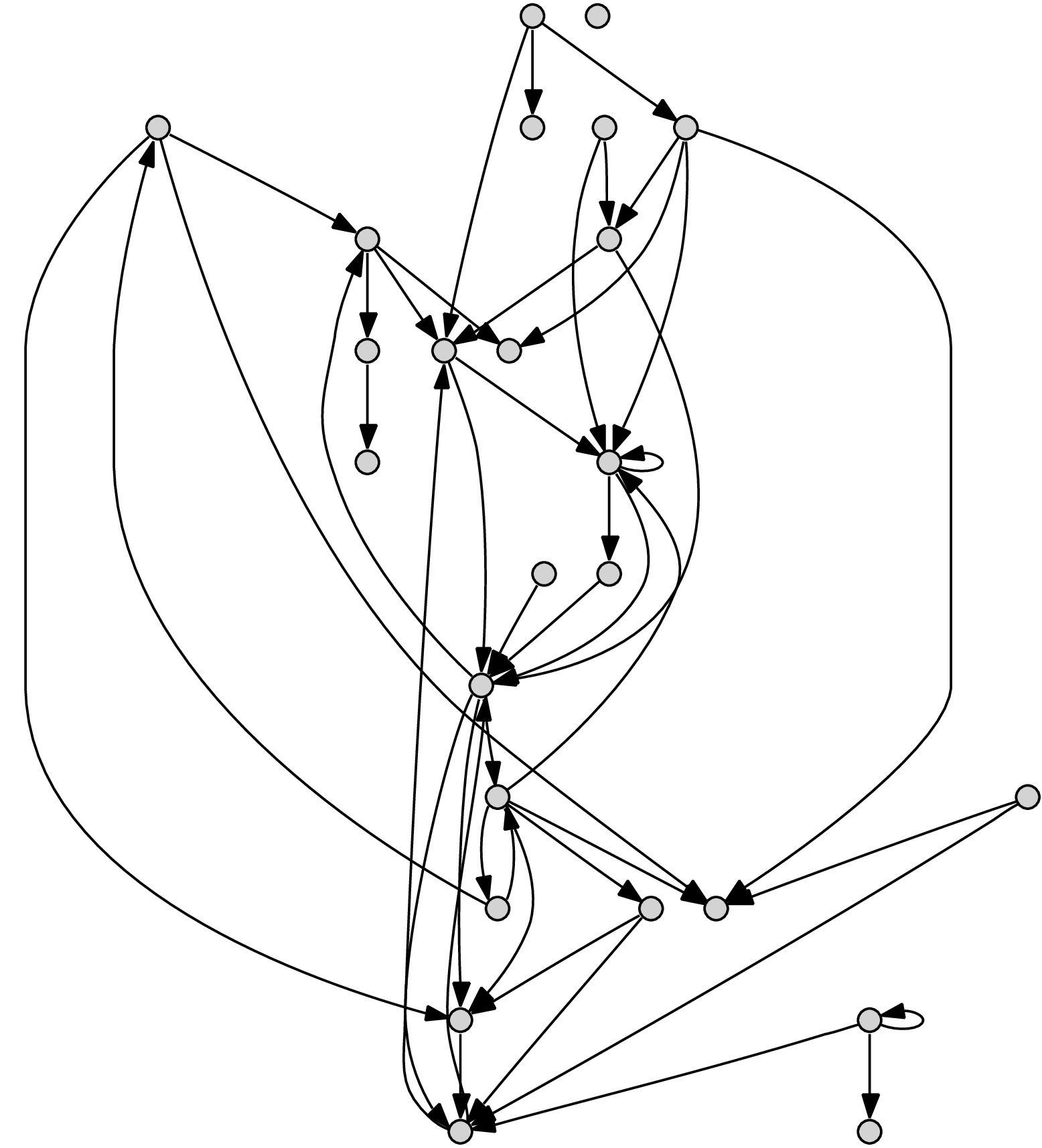} \\
    \hline { neato} &
    \includegraphics[width=1.7cm]{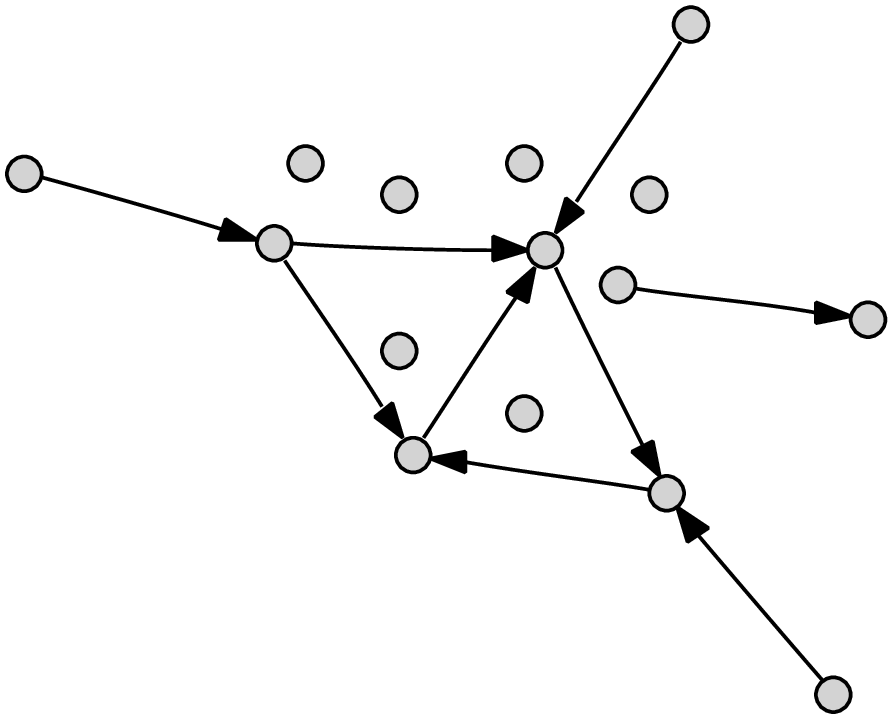} &
    \includegraphics[width=1.7cm]{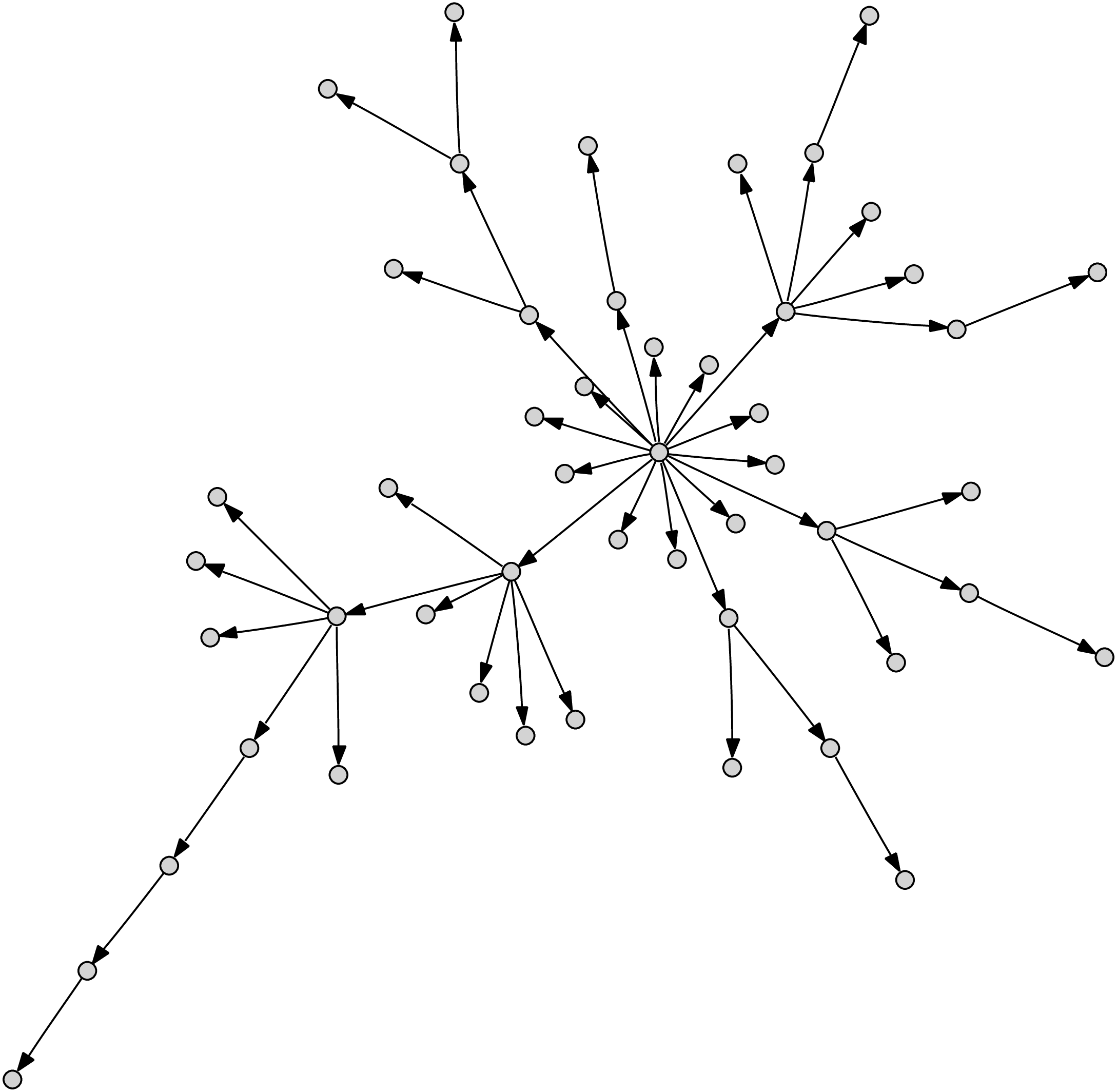} &
    \includegraphics[ width=1.7cm]{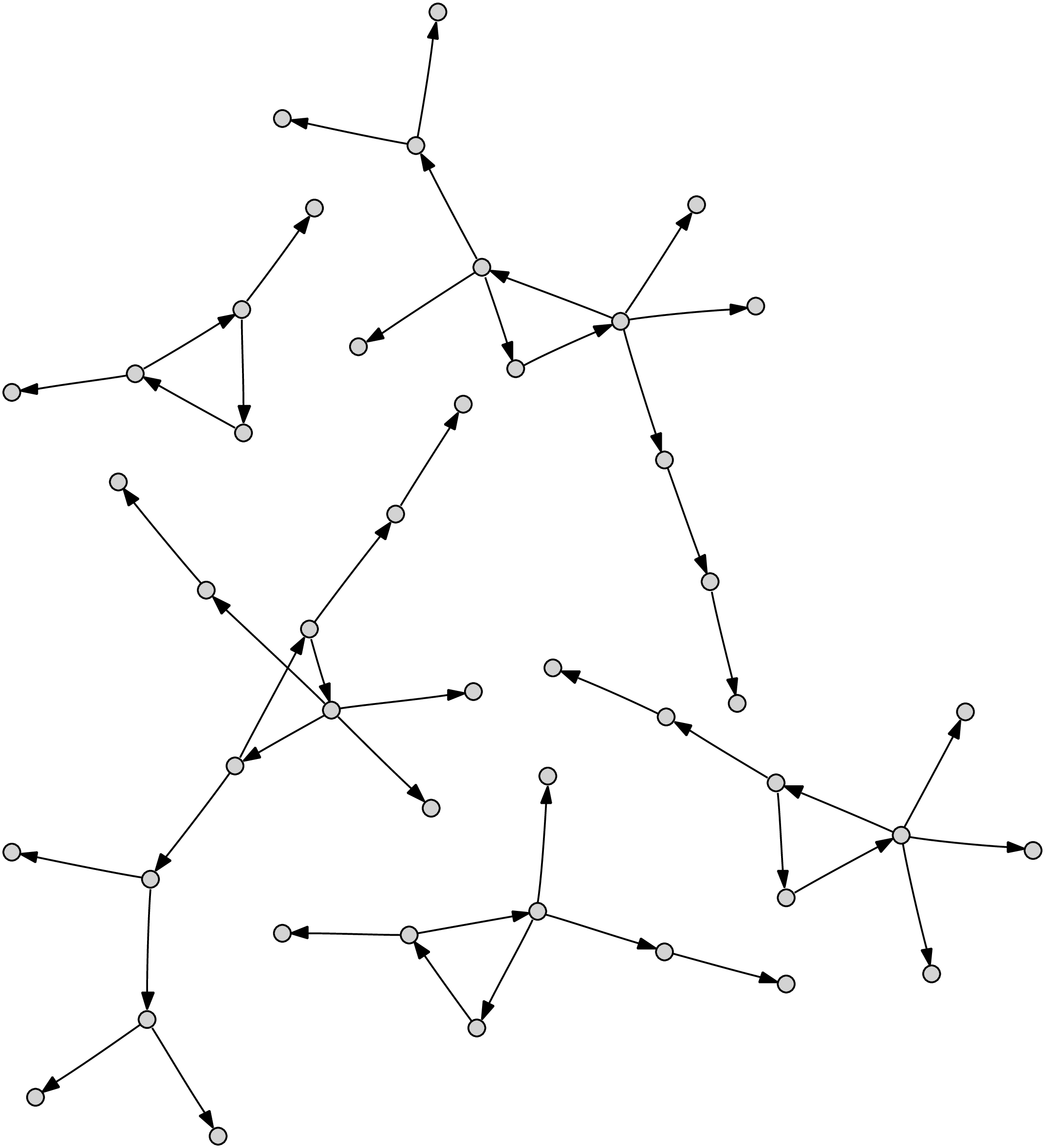} &
    \includegraphics[height=1.7cm]{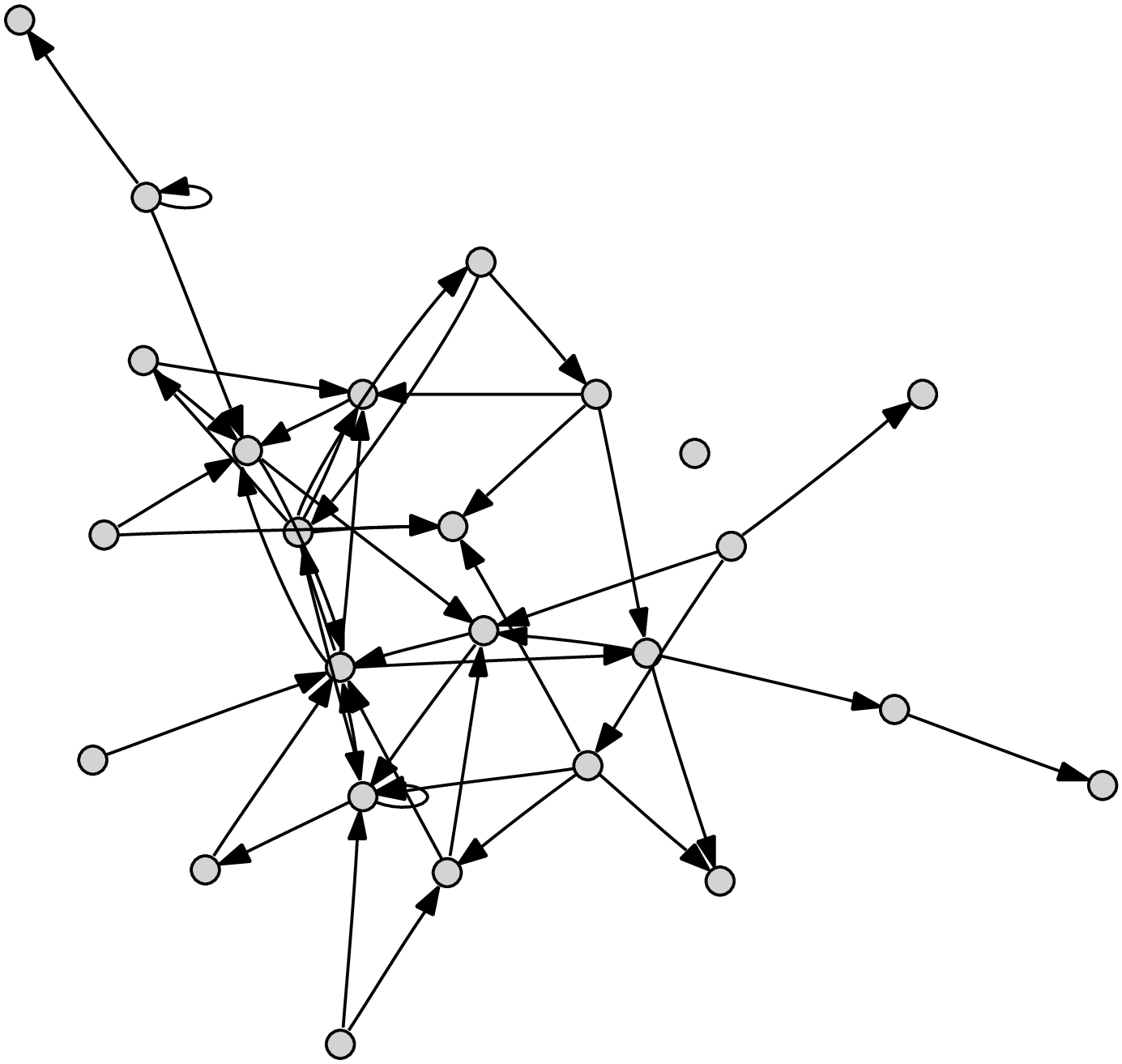} \\
    \hline { circo} &
    \includegraphics[width=1.7cm]{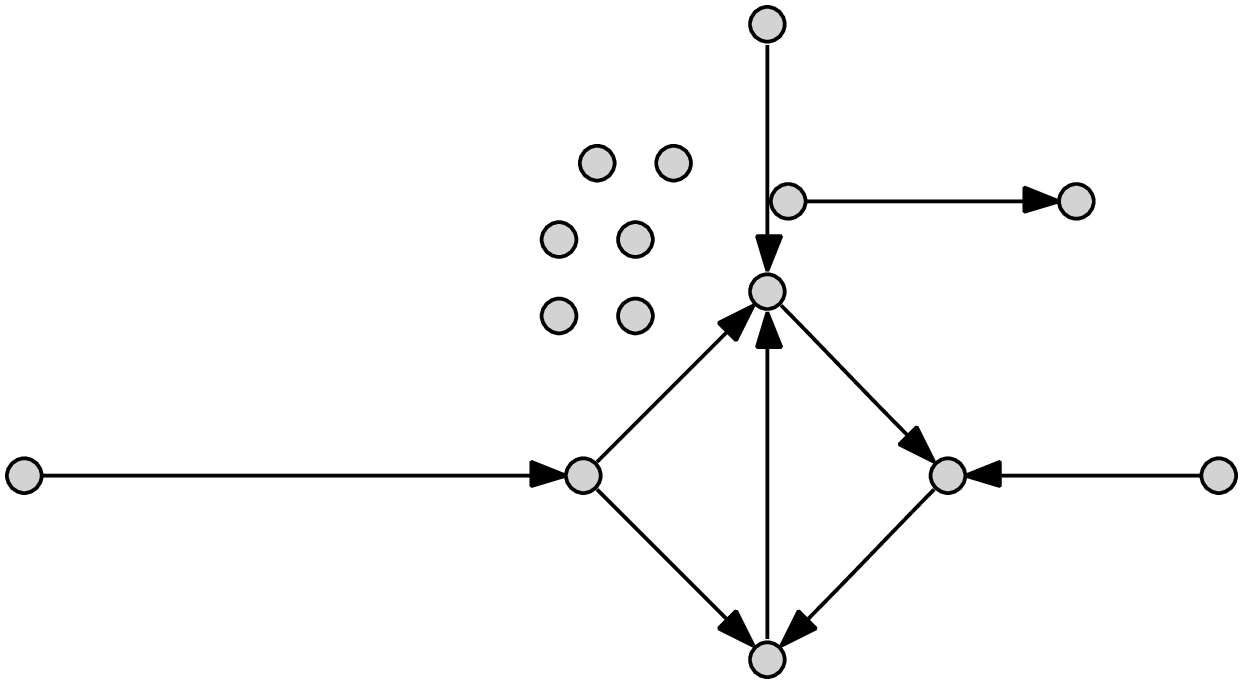} &
    \includegraphics[width=1.7cm]{circo_scalefree} &
    \includegraphics[ width=1.7cm]{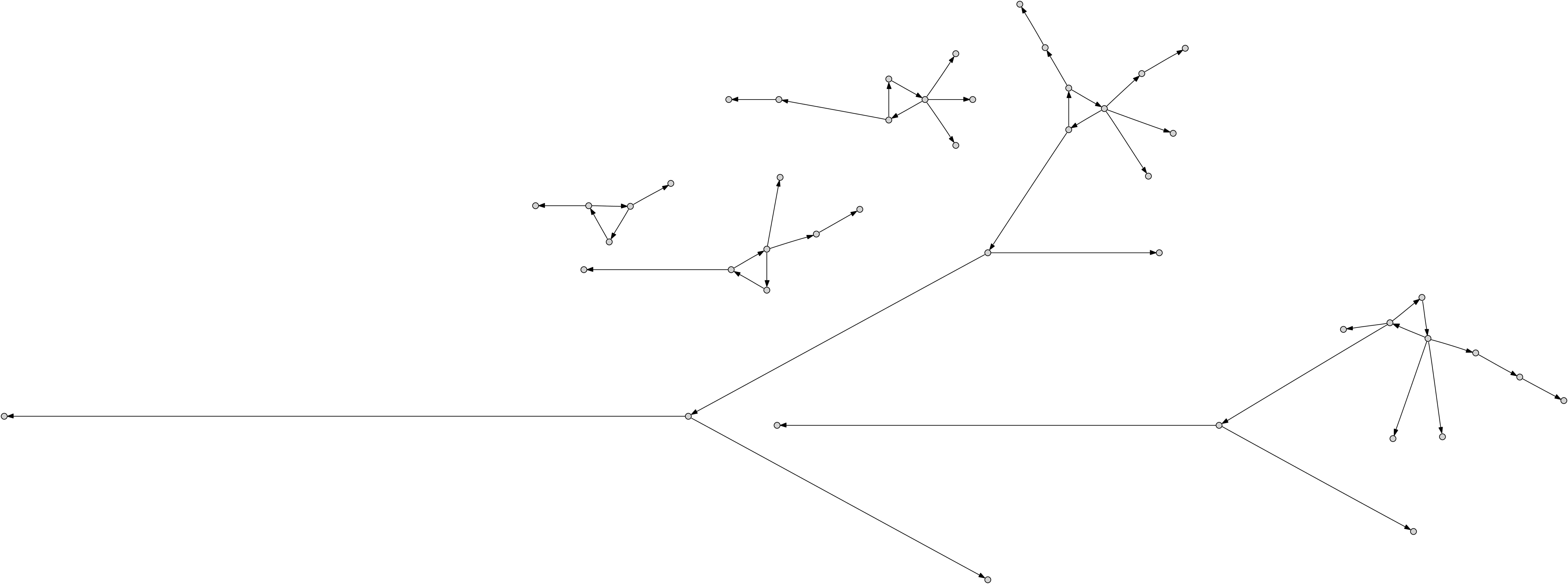} &
    \includegraphics[height=1.7cm]{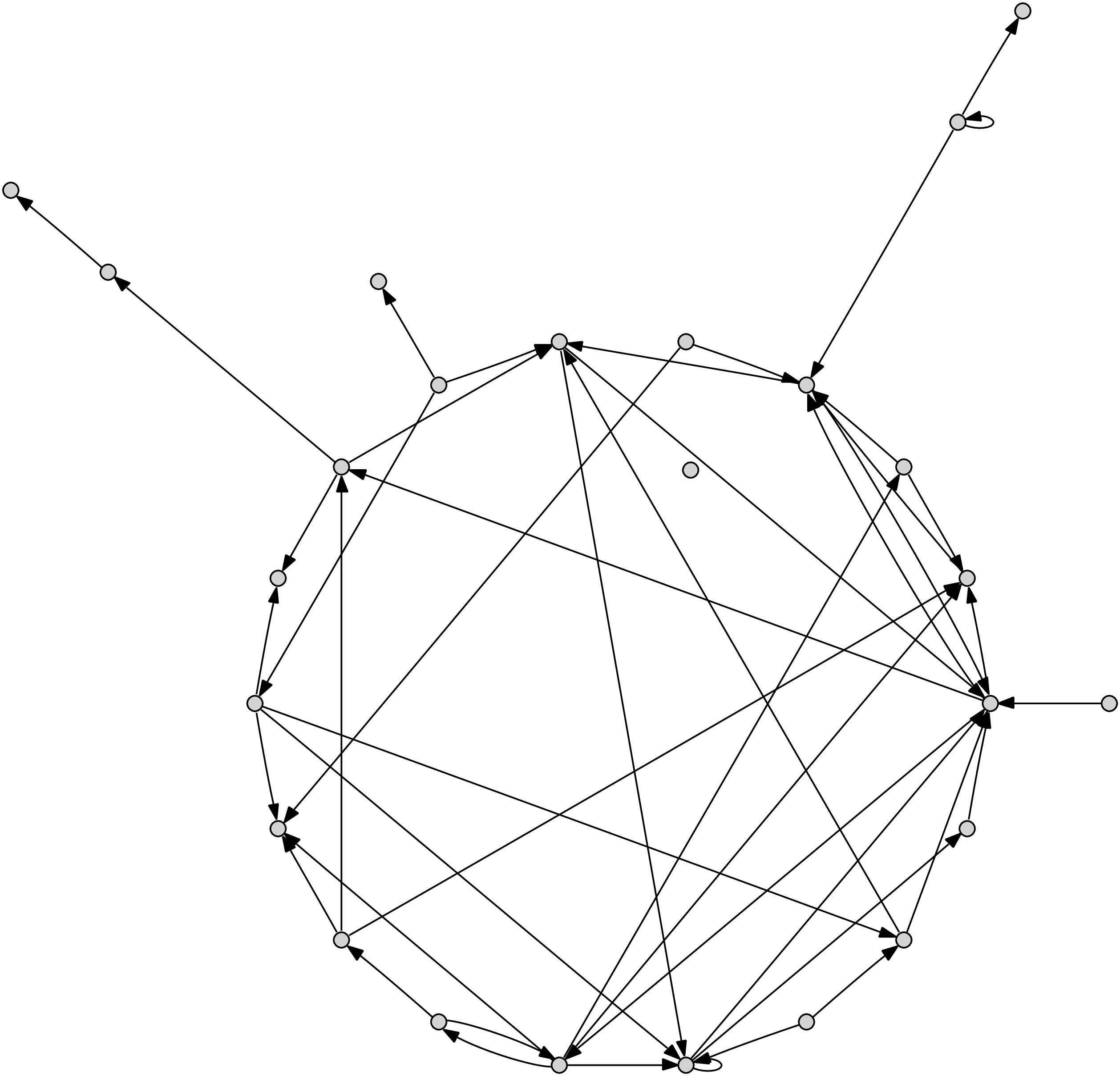} \\
    \hline { twopi} &
    \includegraphics[ width=1.7cm]{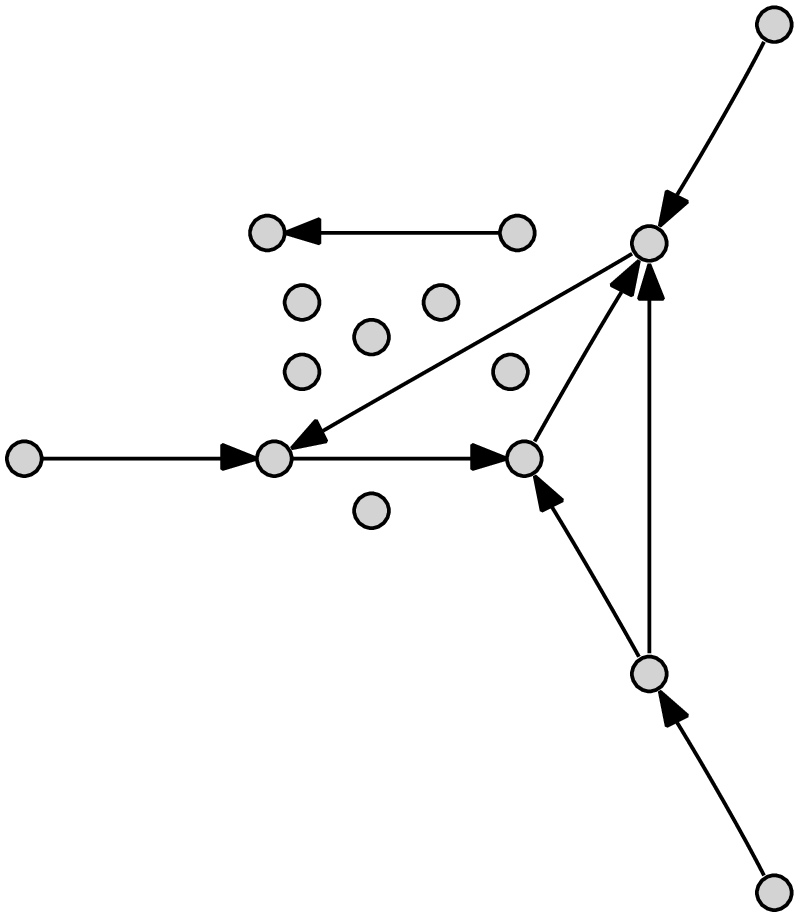} &
    \includegraphics[width=1.7cm]{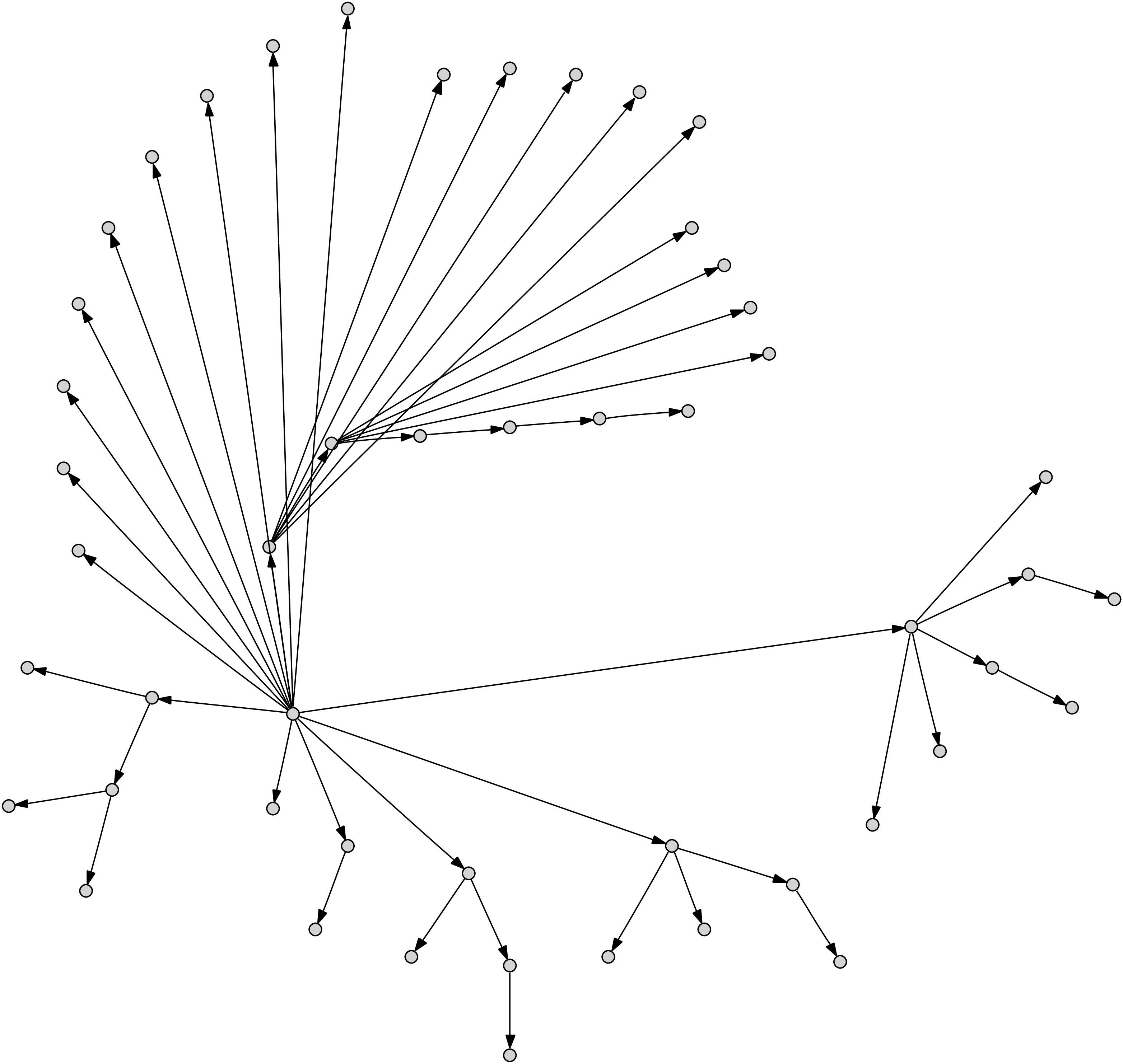} &
    \includegraphics[ width=1.7cm]{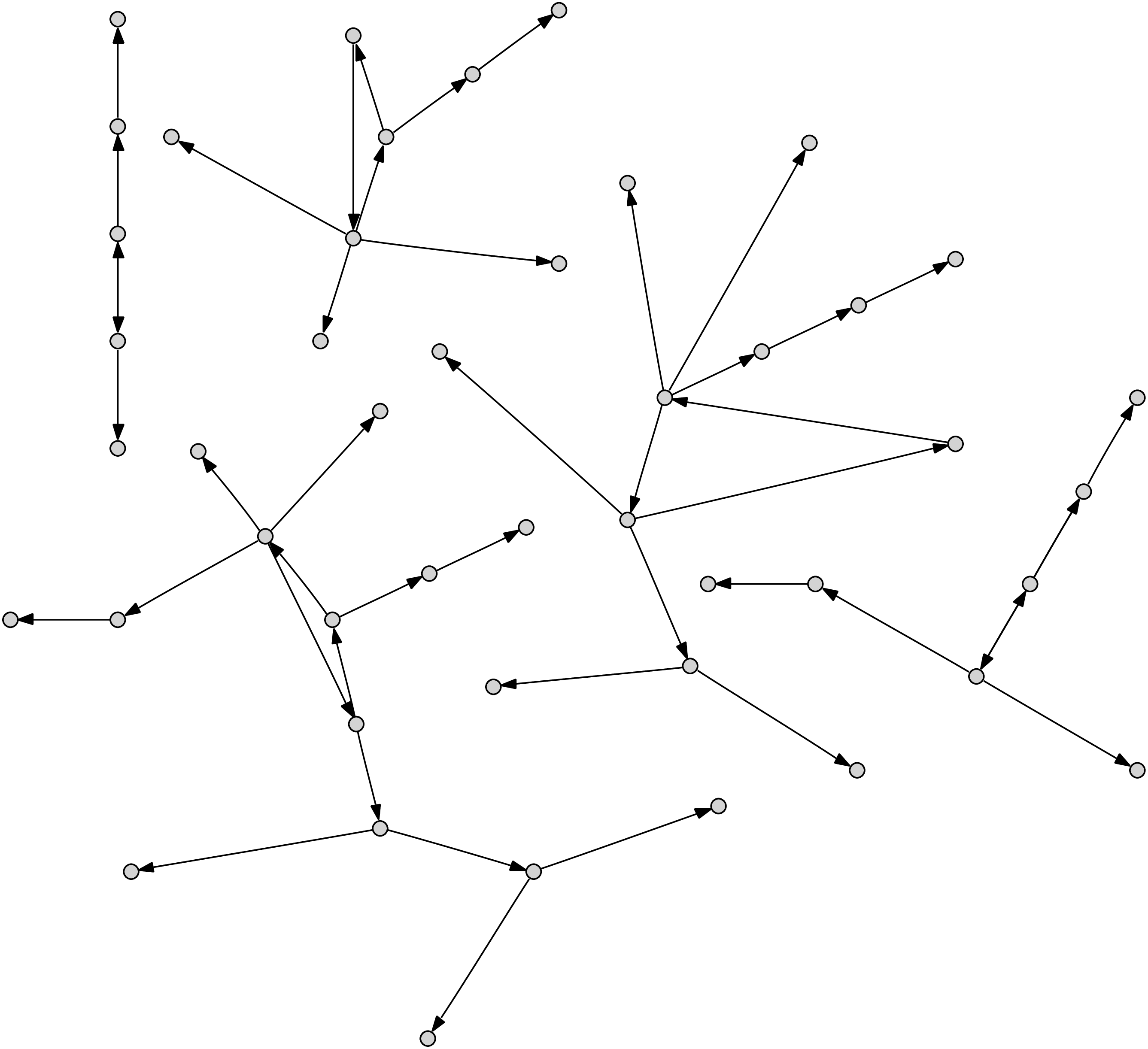} &
    \includegraphics[height=1.7cm]{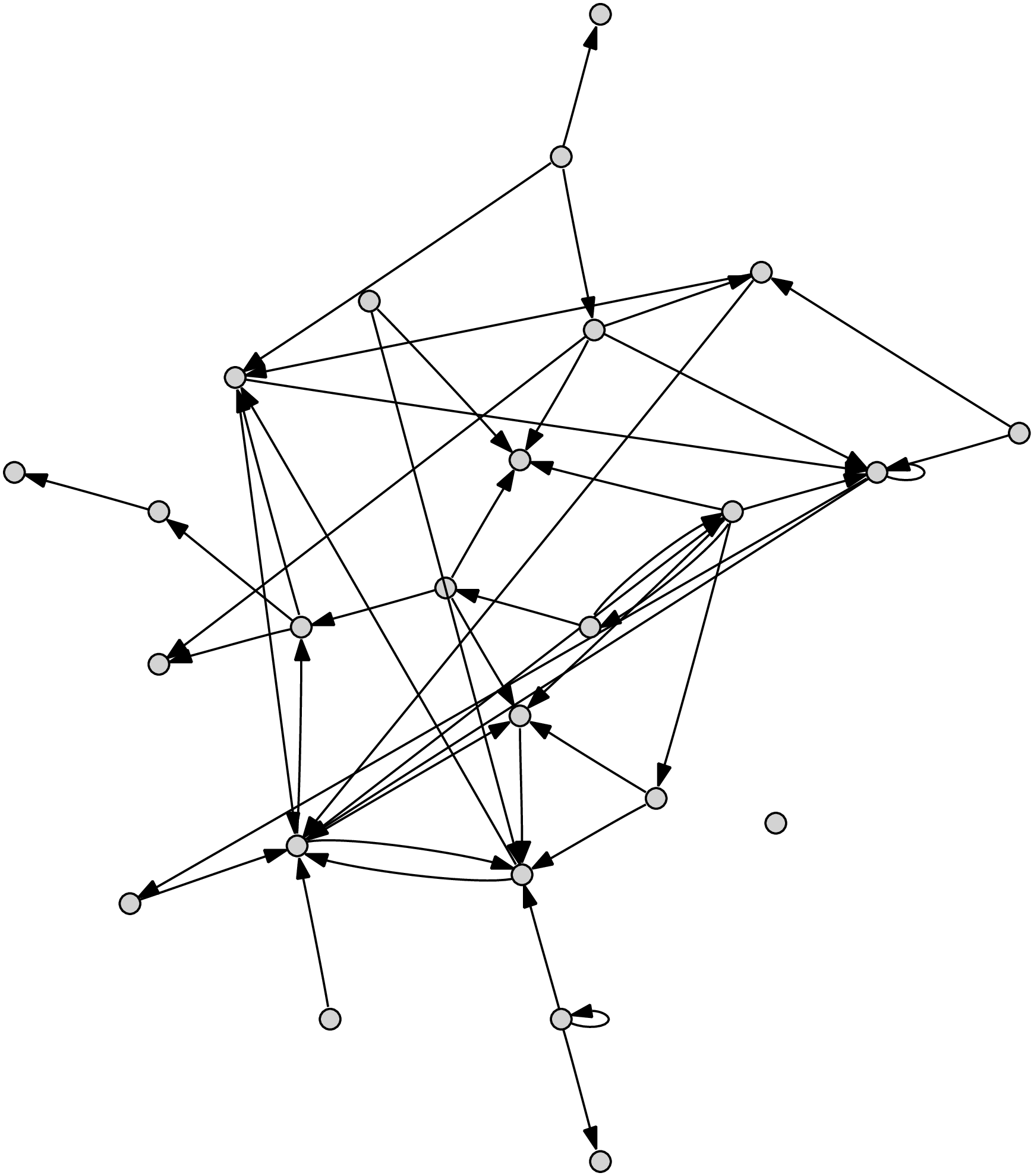} \\
    \hline { fdp} &
    \includegraphics[width=1.7cm]{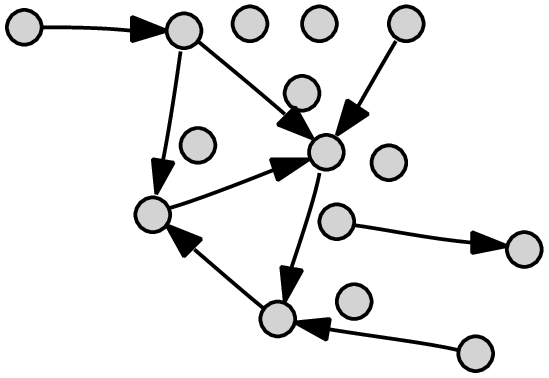} &
    \includegraphics[width=1.7cm]{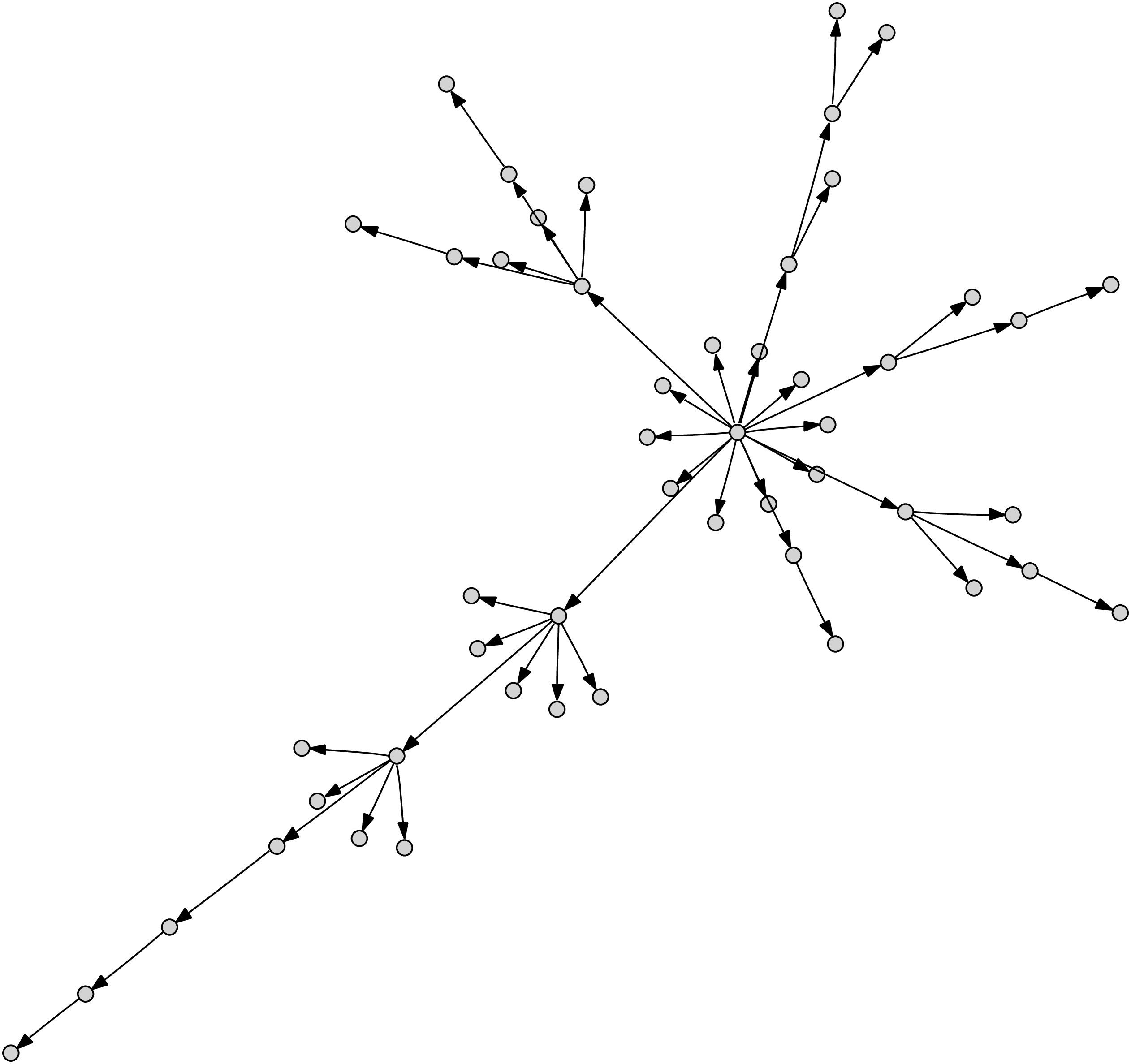} &
    \includegraphics[ width=1.7cm]{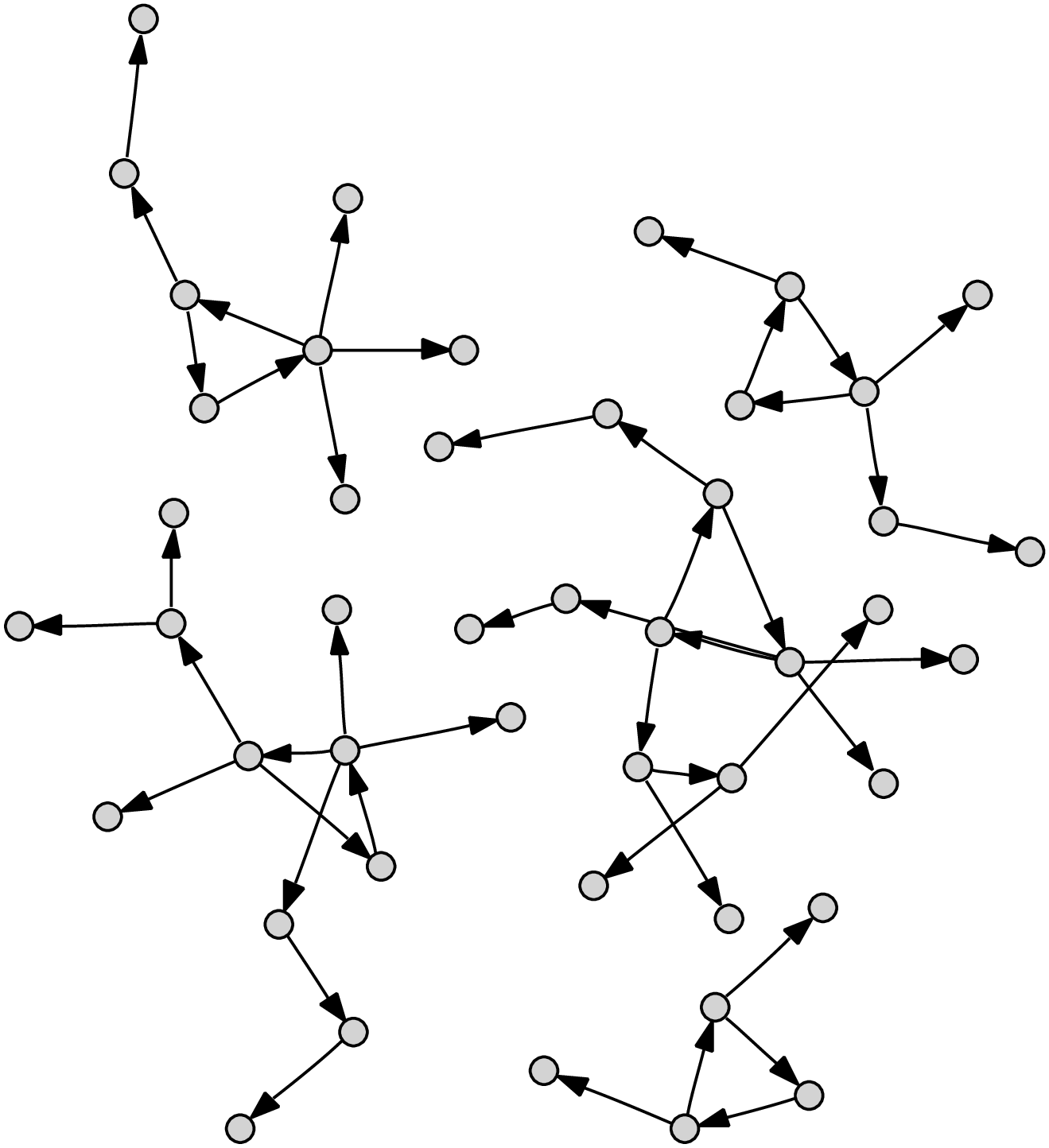} &
    \includegraphics[width=1.7cm]{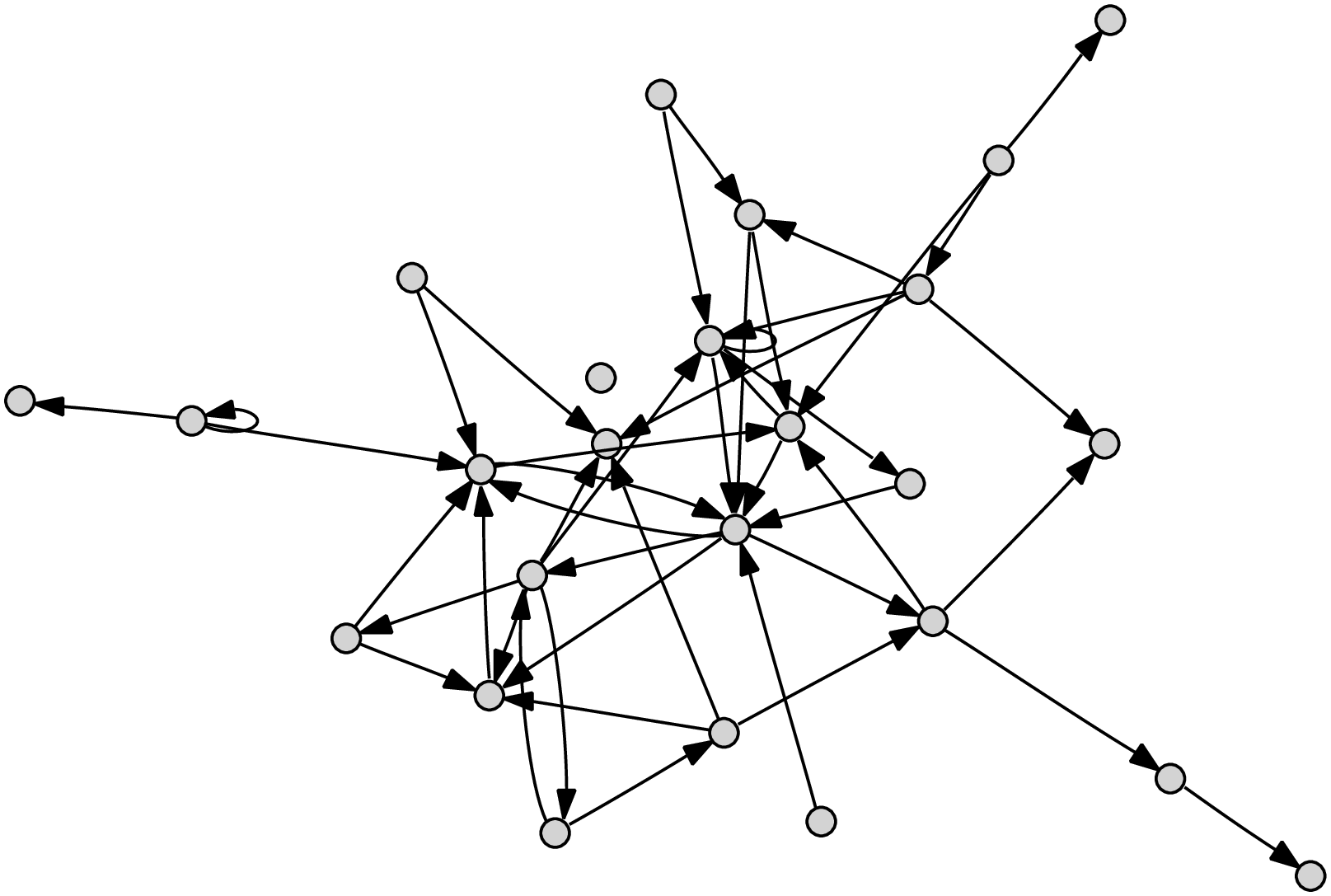} \\
    \hline
  \end{tabular}
 
  \caption{Four different networks where visualized by various layouters:
    Arf, dot, neato, circo, twopi and fdp. The graphs can be
    characterized as follows: E is a network with two connected
    components and many unconnected nodes. F shows a scalefree network.
    G is a set of dentritic clusters. Finally H shows a random
    graph.}
  \label{fig:simpleoverview}
\end{figure}

To address the points ``separation of clusters'' and ``usage of the
available space'', figure \ref{fig:simpleoverview} shows an overview of
layouts produced by {\em arf} and the {\em graphviz} layouters.  The
layouters were all called with the same input dot-file and without any
further options\footnote{ To make the results reproducible, the dot-files
  are available at \url{http://www.sg.ethz.ch/research/} as is a
  demonstrator java-applet for {\em arf}, including the source code.}.
The first row shows the layout produced by {\em arf}.  The other ones
show the layouts produced by the different layouters in the {\em
  graphviz} package.
Right for the first graph (E), an advantage of {\em arf} can be seen:
connected subgraphs are clearly separated, whereas in the layout produced
by {\em neato}, {\em twopi}, and {\em fdp}, single nodes get ``caught''
in the ``kite''-like subgraph.  Looking at the layouts of graph (F) - a
scale free network - the next column clearly shows {\em arf}'s ability to use the
layout space economically can clearly be seen.  Also for the dentritic
clusters in graph (G), {\em arf} produces competitive results: The
clusters are clearly separated and crossing free.  Finally, {\em arf}
produces a clearer and thus more readable layout for a random network
(H).  The nodes are more clearly arranged.

\section{Application Examples}

This section is devoted to the practical application of {\em arf} in
the field of complex systems research.  The first examples deals with the
visualization of social network diagrams and auto catalytic networks.
Next we show how state of the art animations can be generated in
order to shed light on the dynamics of networks with changing structure.
Finally, we highlight how the {\em arf} model can be fruitfully
employed to explore large scale networks.

\subsection{Autocatalytic and Social Networks}

Crucial for interpreting networks is visualization that makes
significant features easy to be perceived by the human eye. The following
examples illustrate this.

\begin{figure}[htb]
   \centering
   \begin{tabular}{c c}
     \includegraphics[height=0.4\textwidth]{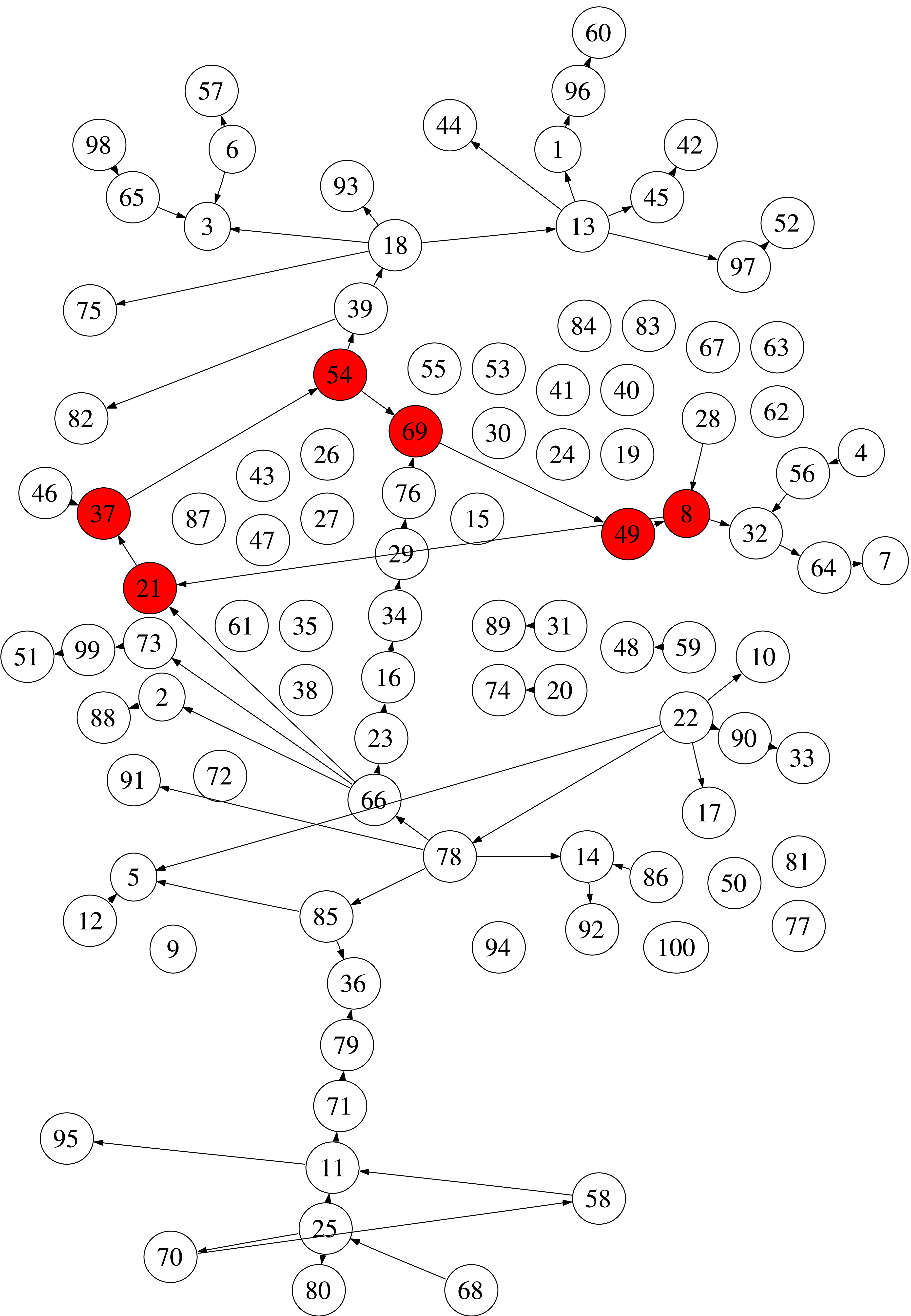}&
     \includegraphics[ width=0.4\textwidth]{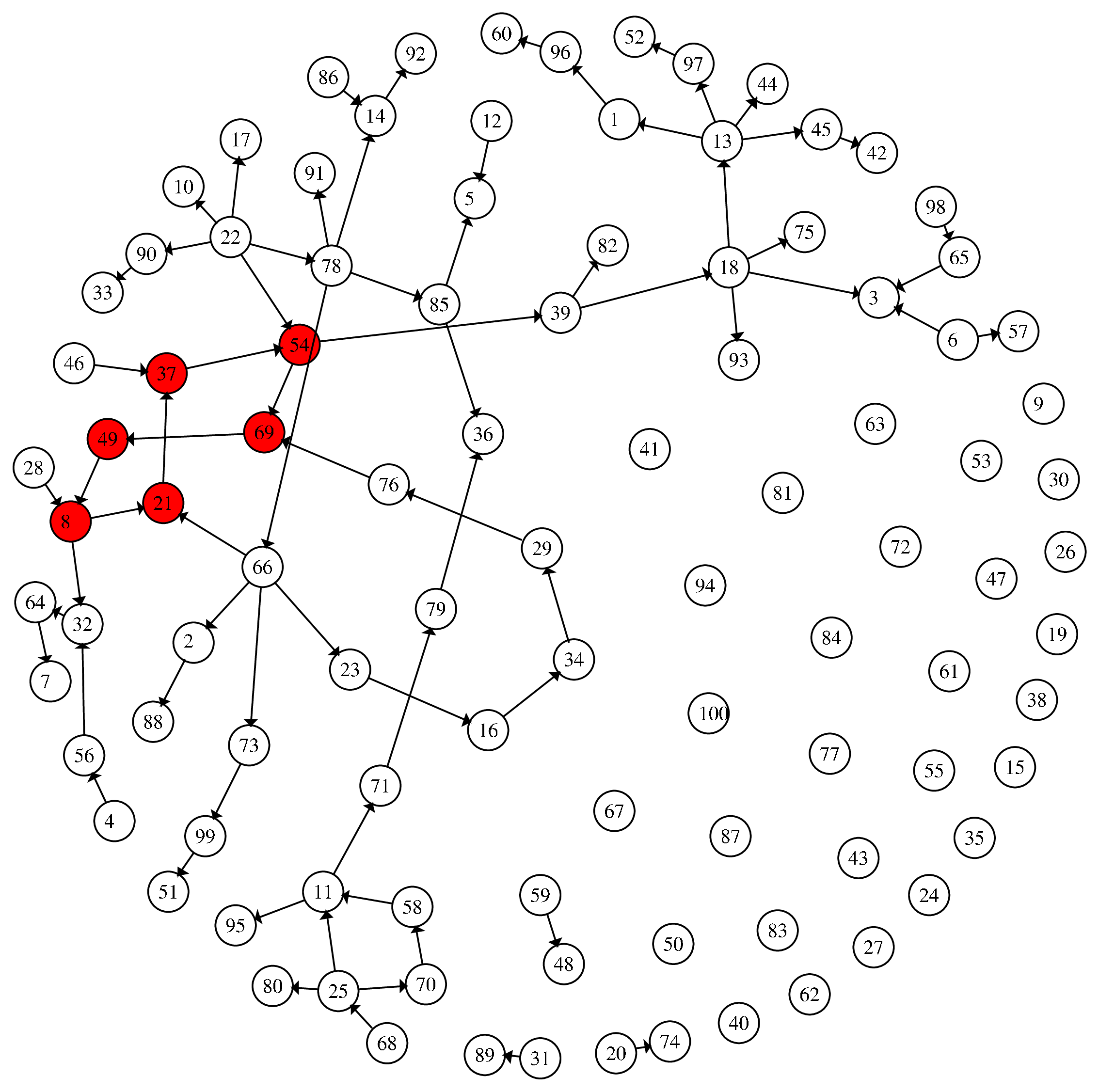}\\
     a & b\\
   \end{tabular}
   \caption{An auto catalytic network taken from \cite{seufert06}, laid
     out by {\em twopi} (a) and {\em arf} (b). {\em Arf} clearly shows
     the structural features of the graph: one big connected component
     with a dentritic periphery and a six-cycle (marked red).}
   \label{fig:seufert}
 \end{figure}

Figure \ref{fig:seufert} shows an autocatalytic network, taken from
\cite{seufert06}.  The network was laid out twice: The layout on the left
is taken from the original publication (it was generated with {\em
  twopi}). The layout produced by {\em arf} (right) is far easier to read
and to interpret: The networks contains one big connected component,
three two-node-components and numerous single nodes.  The connected
component can be divided into a core forming an autocatalytic cycle
(marked red) and a dentritic periphery.  All these characteristic
features are clearly visible in the right layout but not in the
left one.

 Next we revisit the social network presented in section \ref{sec:intro}.
 In figure \ref{fig:manager} managers of firms are connected by joint
 board membership\footnote{The data is taken
   from the ORBIS 07 database (Bureau van Dijk).}.  Sub figure
 \ref{fig:manager} a and b are generated with {\em neato} and {\em fdp}.
 Sub figure \ref{fig:manager} c was generated with {\em arf}. Even though
 the big star takes more space in c than in a and b, both the strongly
 connected core, as well as the periphery are given more space than in
 the other visualizations. Their structure is clearer and it is easier to
 discern different cliques of nodes.

 \subsection{Animating Dynamic Graphs}
 \label{sec:dquality}

 In this section we describe how {\em arf} can be used to produce state
 of the art animations.  The basic scheme is shown in algorithm (2).
 \begin{algorithm}[htb]
   \label{alg:anim}
   \caption{Animate($G$)}
   \begin{algorithmic}[1]
     \For {{\bf all} $c \in C$} \State perform $c$ on $G$ \For {$k$ {\bf
         times}} \State relayout $G$ \State render $G$
     \EndFor
     \EndFor
   \end{algorithmic}
 \end{algorithm}
 In in the outer for loop, the algorithm iterates through all the changes
 $c$ in the set of changes $C$ that are to be performed to the graph $G$.
 These changes can, for example, be codified in a script:
\begin{verbatim}
addnode 12, addedge 12 1;
dropedge 12 1;
\end{verbatim}
 A similar script approach was also presented by \cite{brandes00} to
 codify and visualize changes in the WWW.  This example would
 simultaneously add a node with id 12 and an edge from it to node 1.  In
 the next graph change, the very same edge would be dropped.  Now that
 the graph changed, the layout has to be adapted.  This is done in $k$
 steps to ensure a smooth transition from layout to layout.  The smaller
  $k$, the faster the animation.  The relayout procedure is just the
 body of the while-loop (lines 2-7) in Algorithm (1).  After each
 relayout the graph is rendered to the screen or an image file,
 respectively.

\begin{figure}[htb]
   \centering
   \begin{tabular}{|c|c|c|c|c|}
     \hline     
     \includegraphics[height=1.8cm]{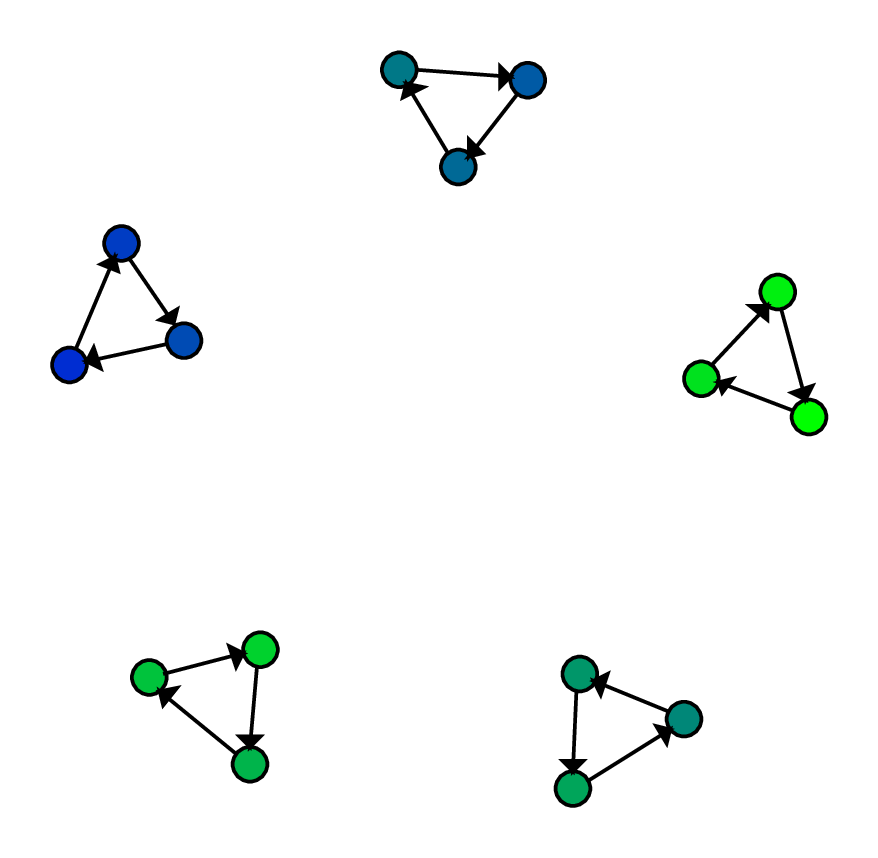}&
     \includegraphics[height=1.8cm]{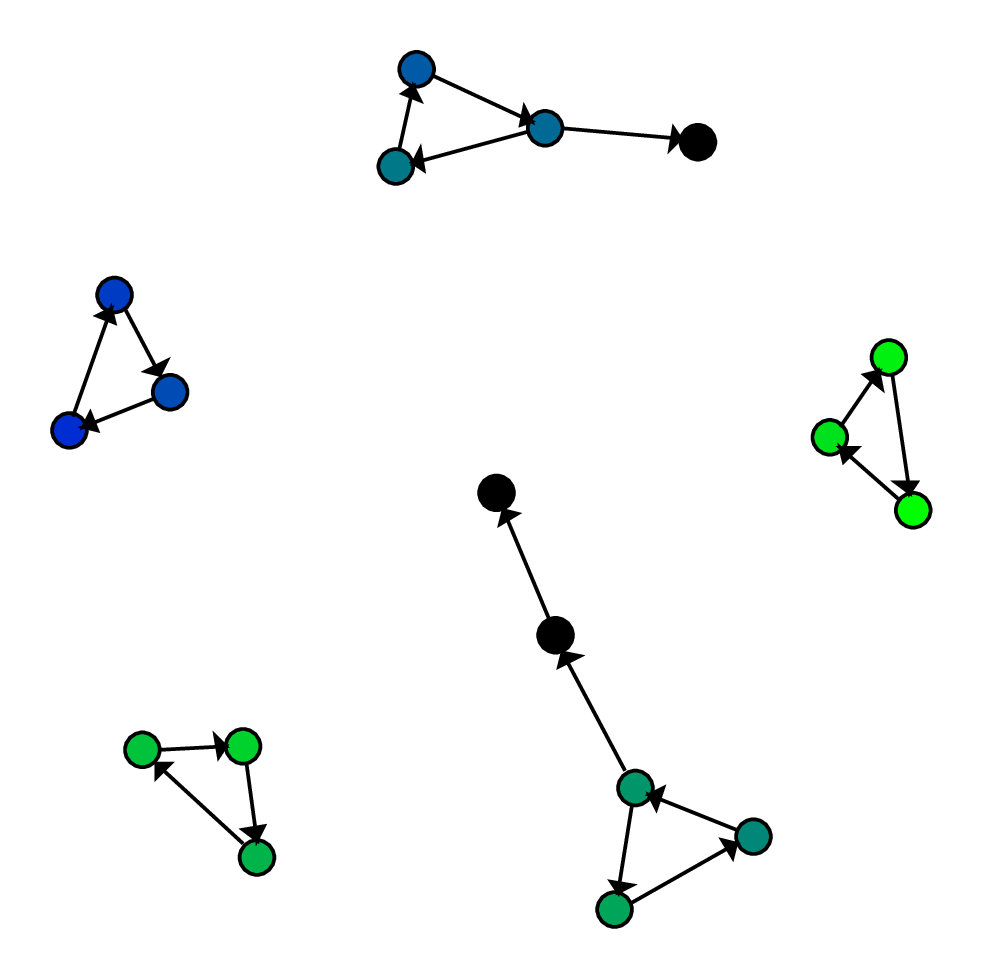}&
     \includegraphics[height=1.8cm]{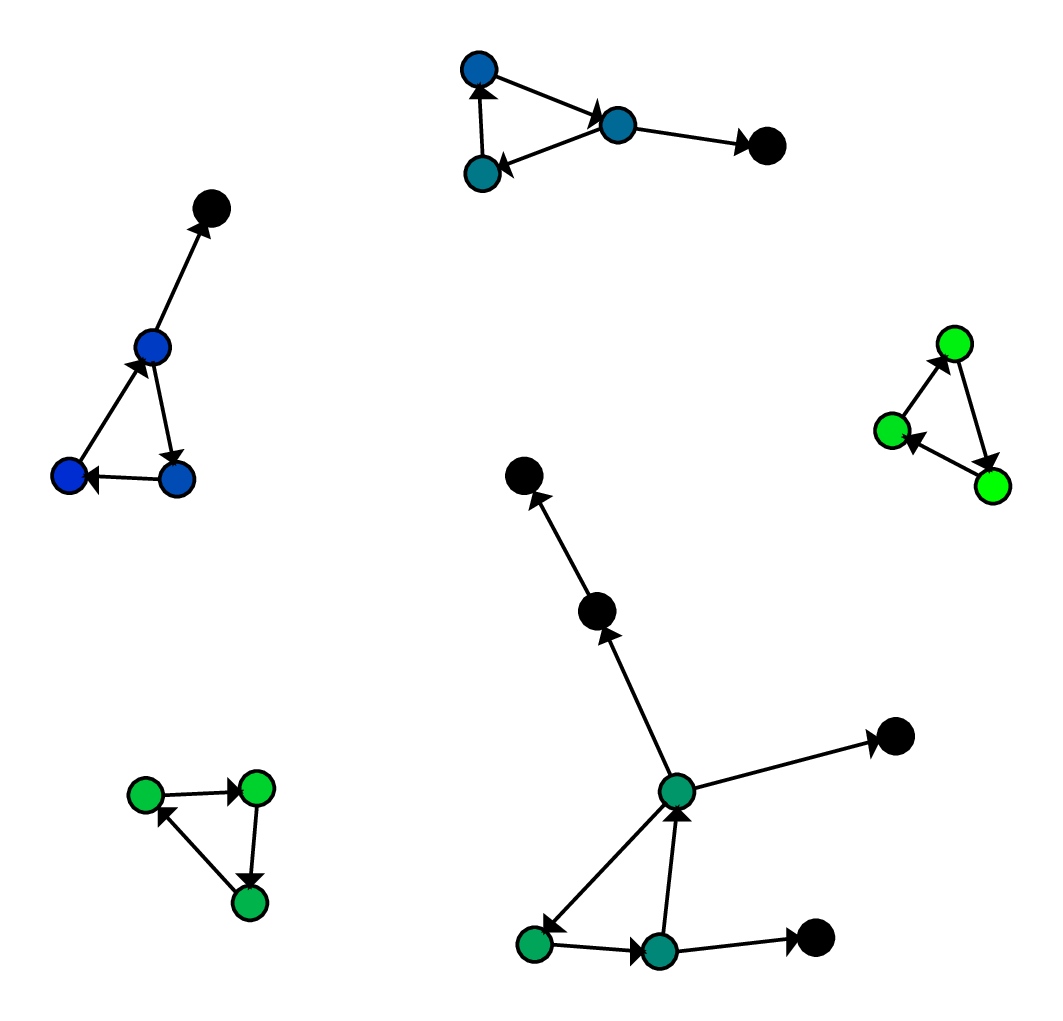}&
     \includegraphics[height=1.8cm]{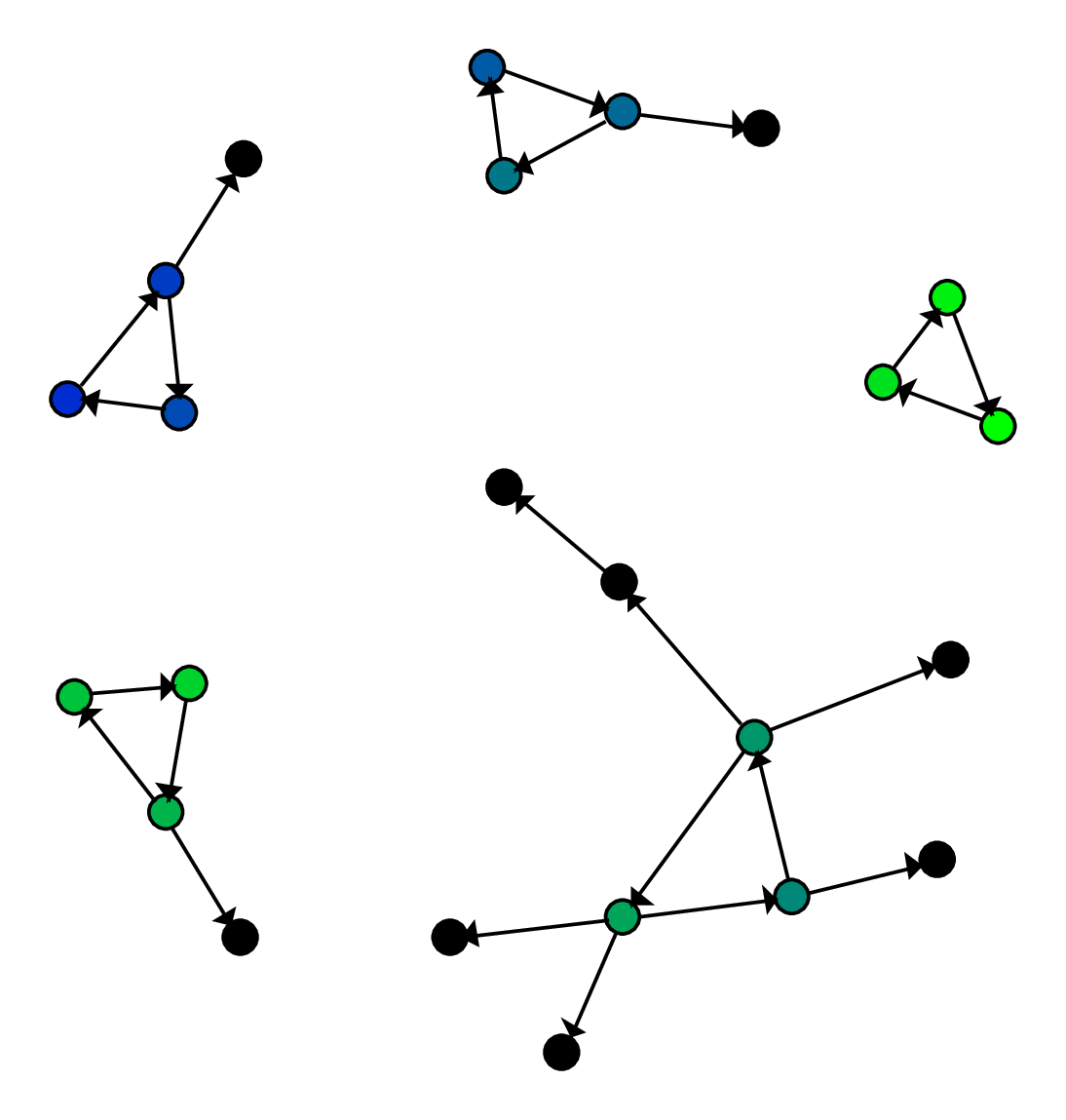}&
     \includegraphics[height=1.8cm]{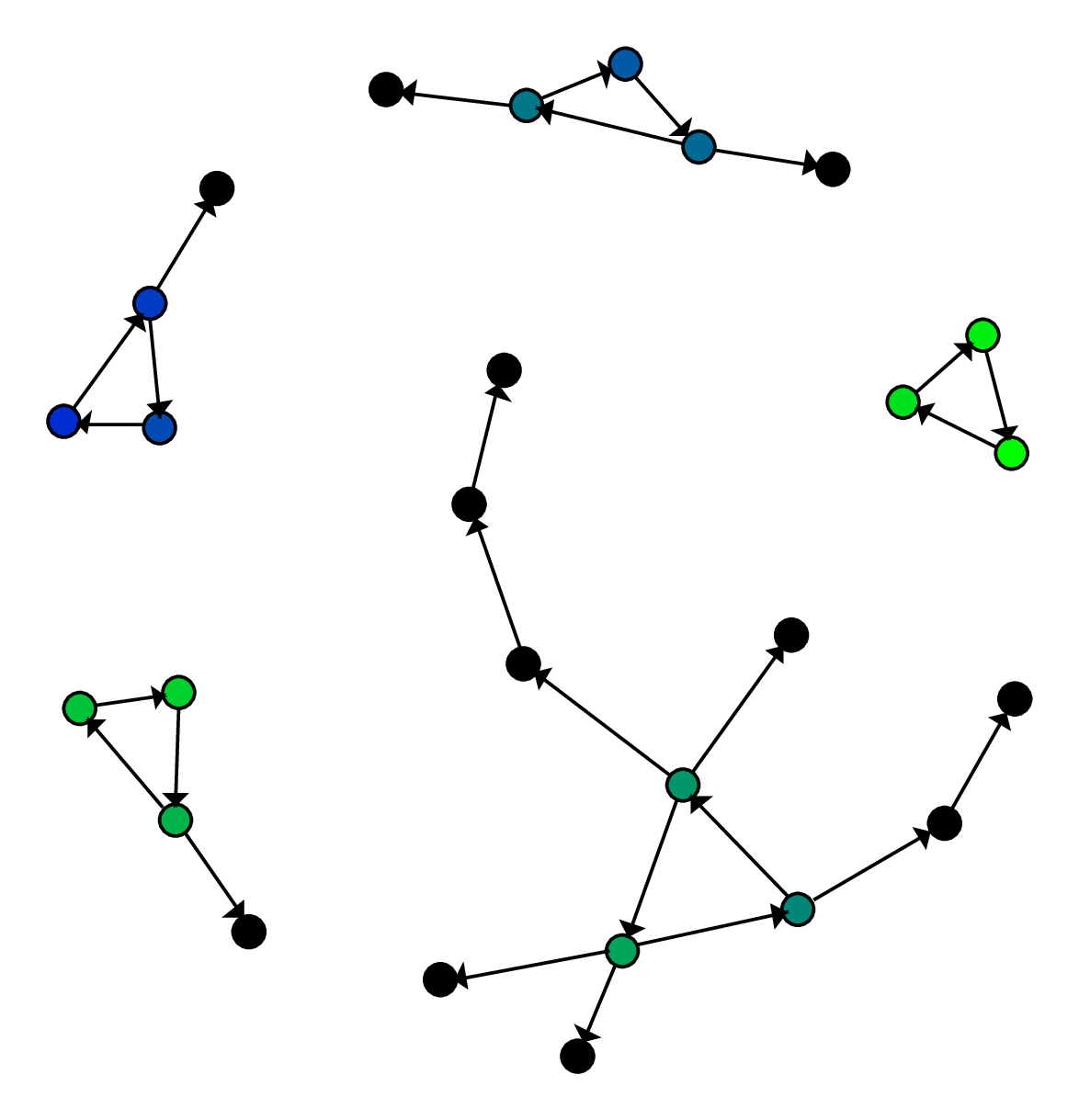}\\
     $t=0 $ & $t=90 $ & $t=180 $ & $t= 270 $ & $t = 360$ \\
     \hline
     \includegraphics[height=1.8cm]{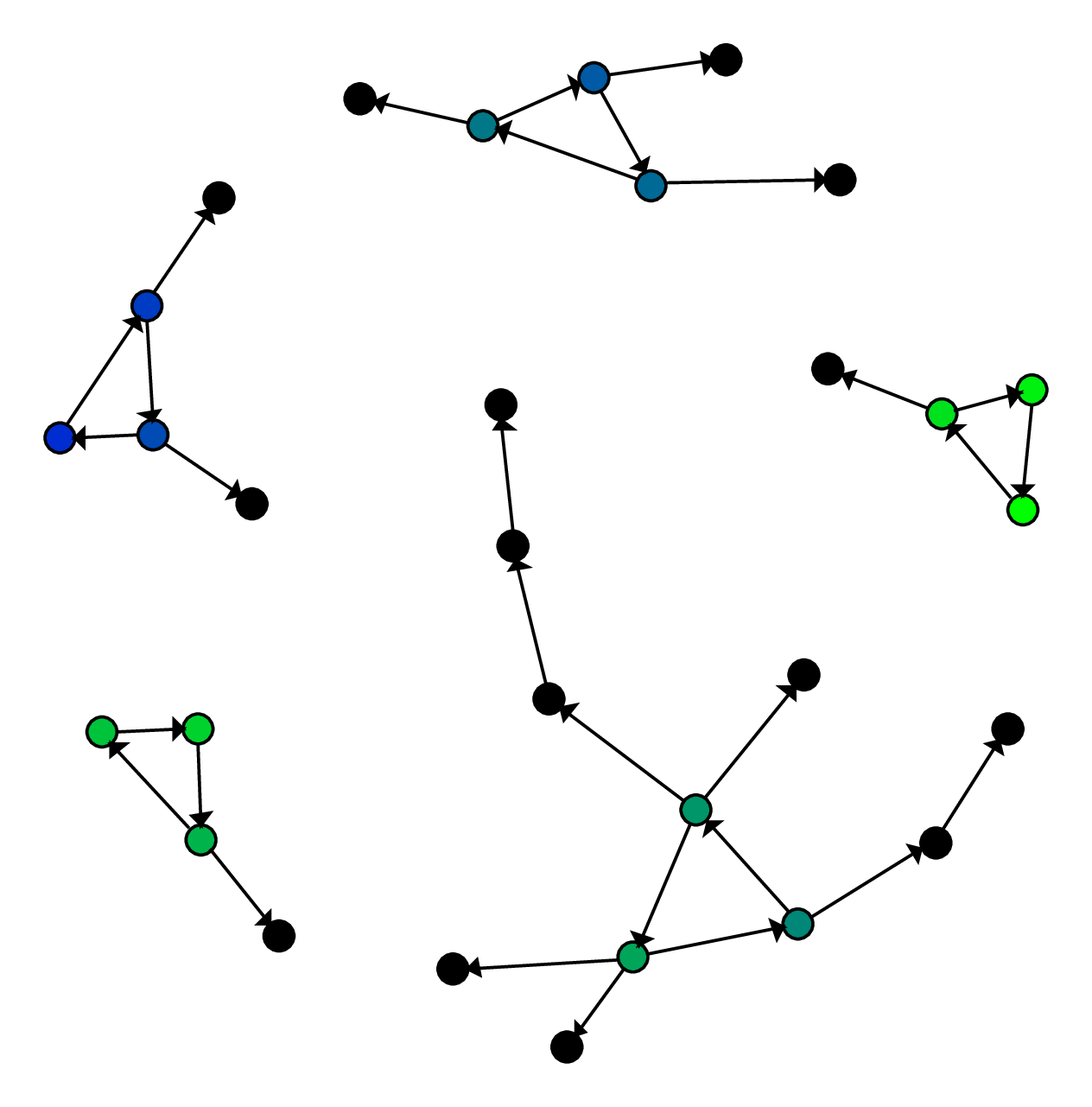}&
     \includegraphics[height=1.8cm]{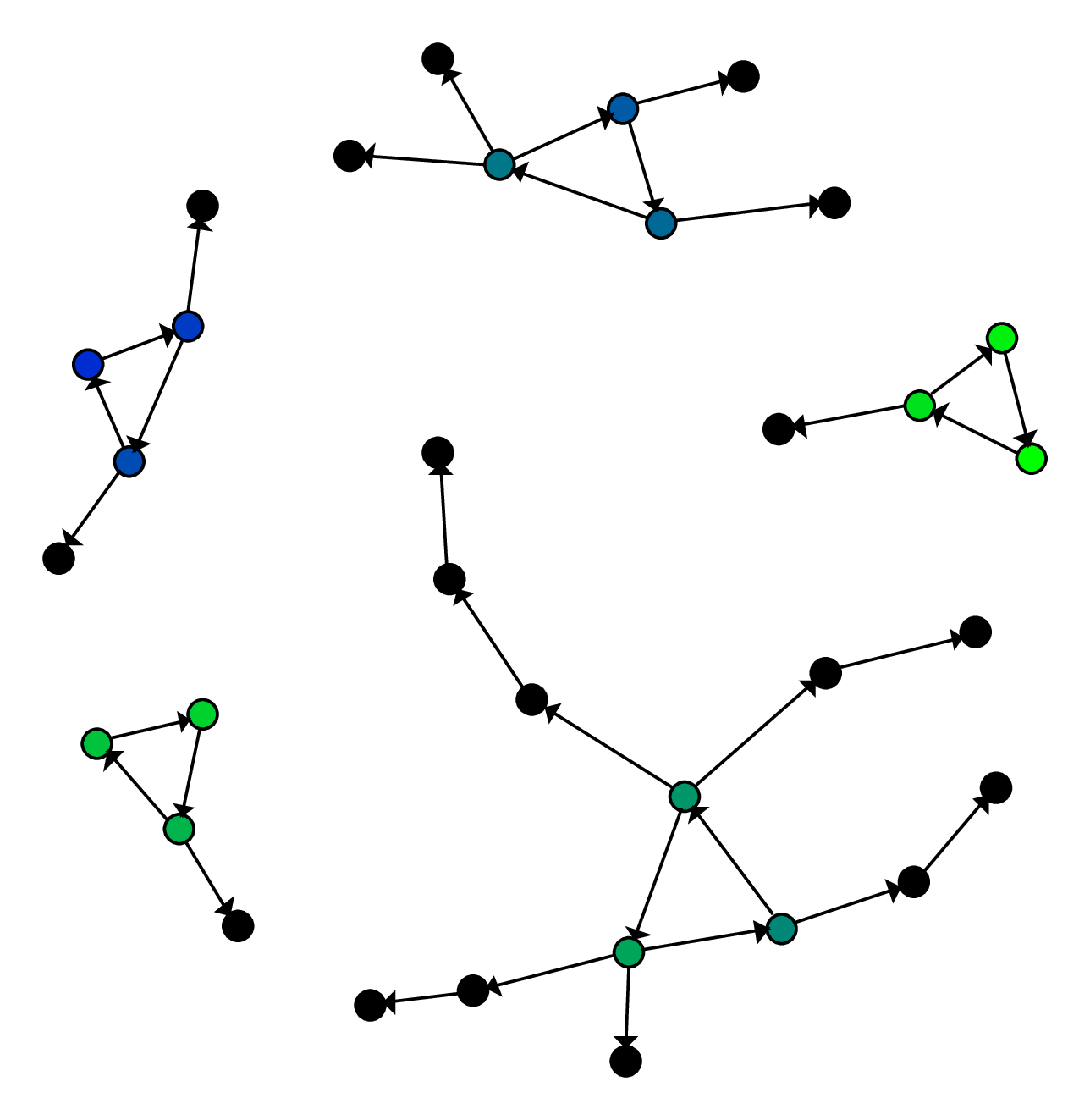}&
     \includegraphics[height=1.8cm]{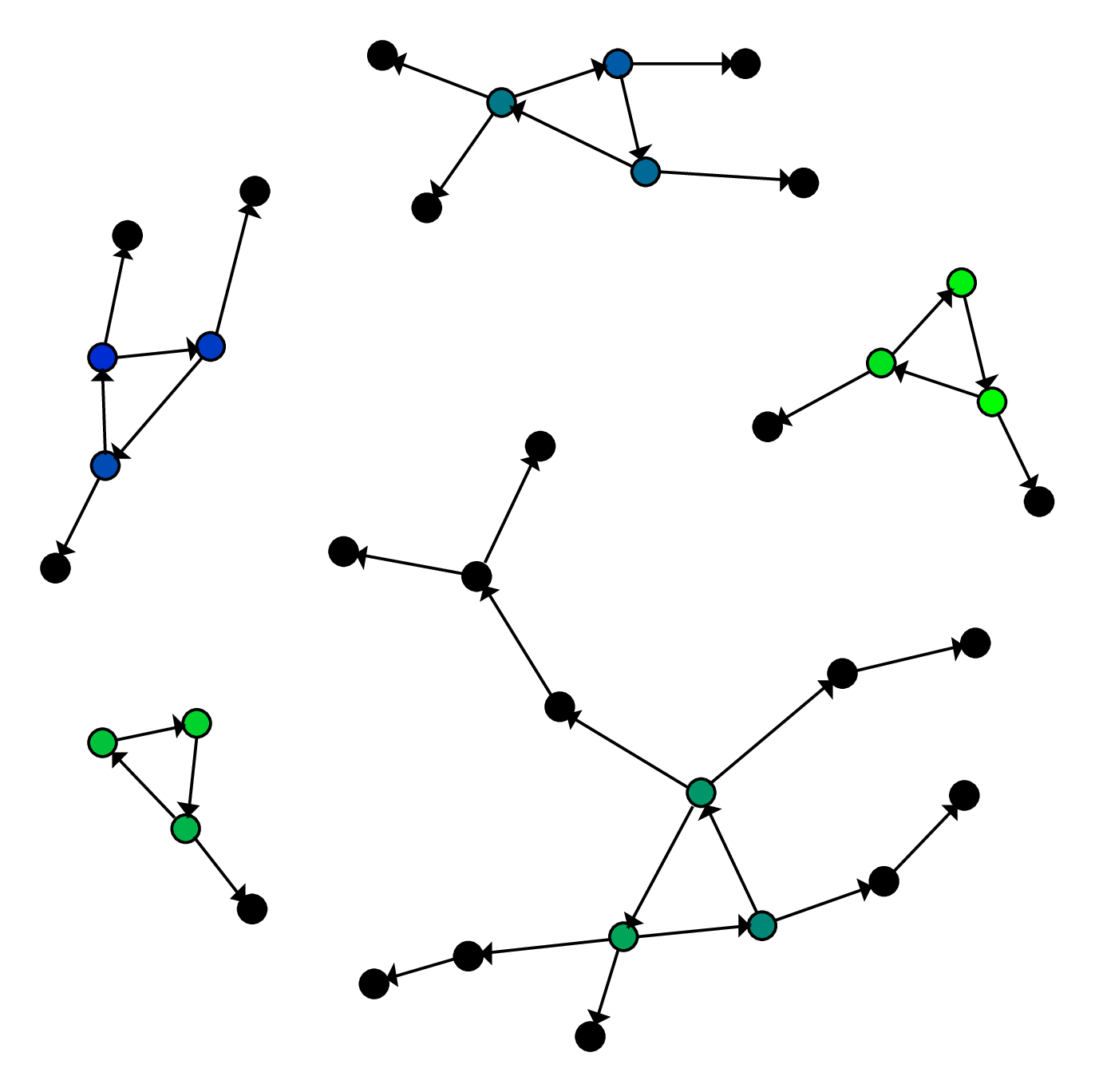}&
     \includegraphics[height=1.8cm]{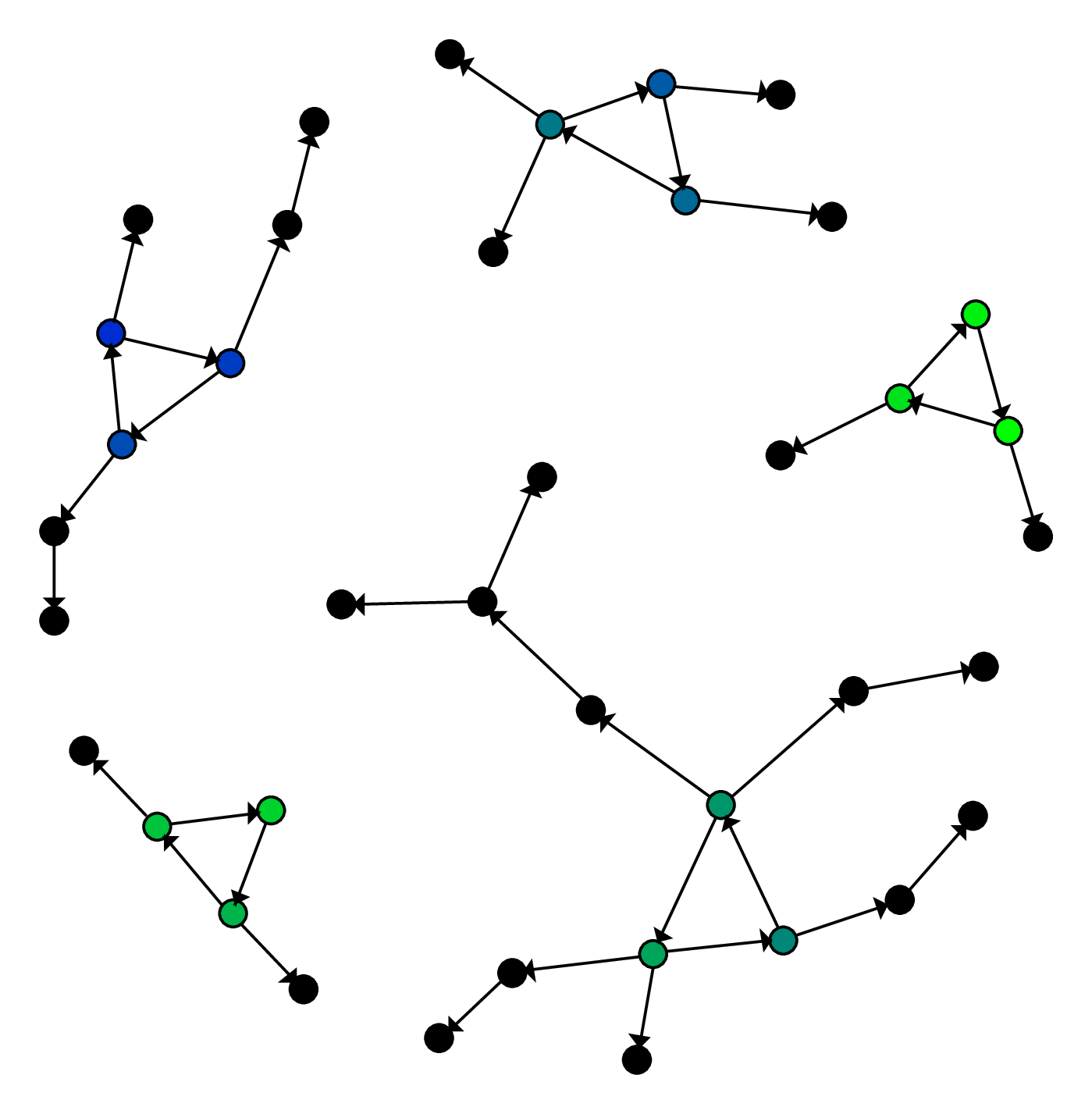}&
     \includegraphics[height=1.8cm]{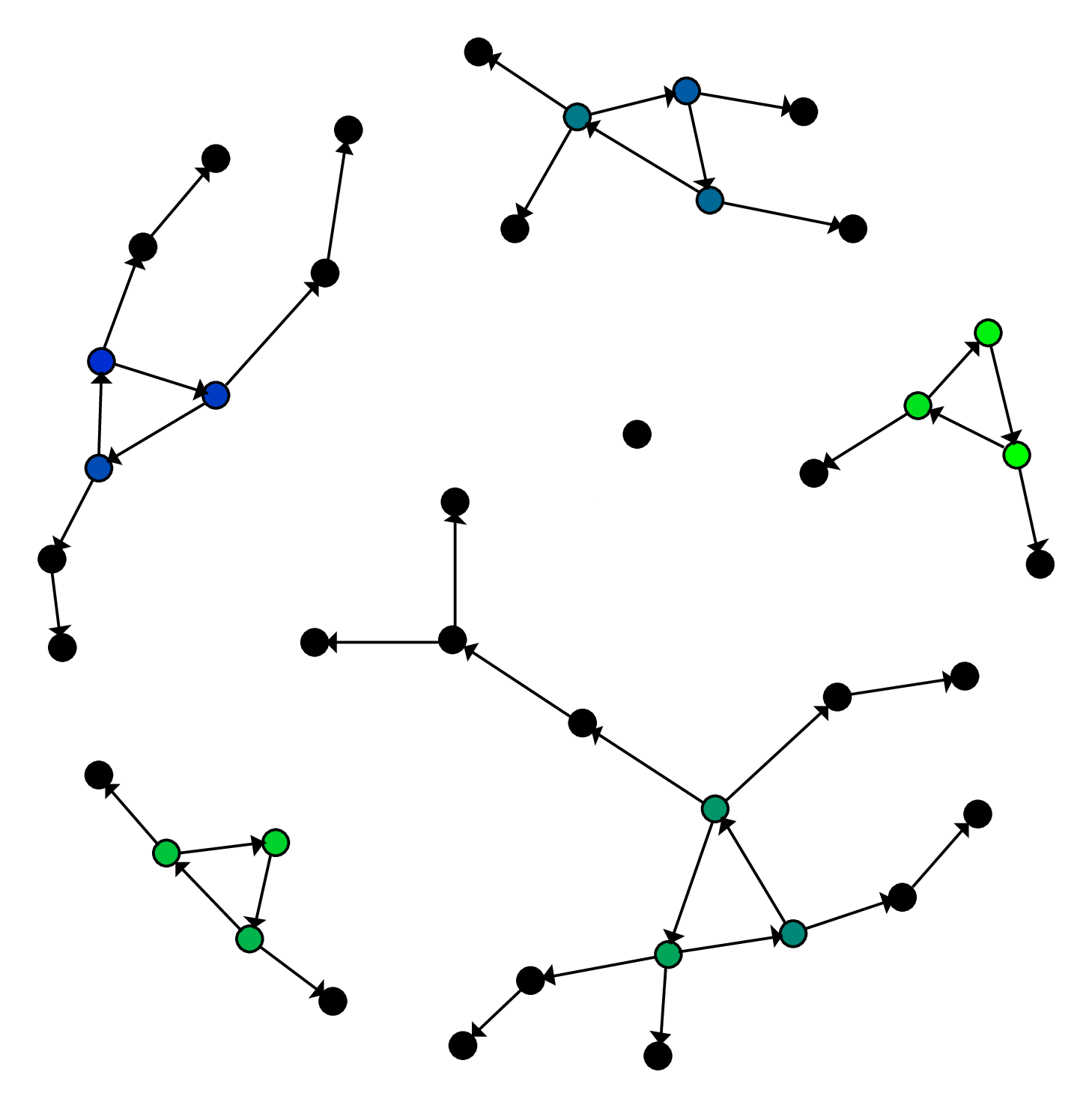}\\
     $t=450 $ & $t=540 $ & $t=630 $ & $t= 720 $ & $t = 810$\\

     \hline

   \end{tabular}
   \caption{A dynamically changing graph laid out with {\em arf}.}
   \label{fig:dynamic}
 \end{figure}

 In the latter case, the generated image files can easily be compiled to
 a video file of any format by using openly available graphics tools.
 Figure \ref{fig:dynamic} shows an example of such an animation.  Exactly
 the same procedure was used in actual research practice: The simulation
 of R\&D cooperations between firms was animated with {\em arf}
 \cite{koenig06}. The videos can be downloaded at
 \url{http://www.sg.ethz.ch/research/graphlayout/} as well as several
 other vi\-deos of evolving networks.

 \subsection{Exploring Large Networks}

 Often networks are too large to be visualized completely. Just imagine
 cross-investment networks spanning entire countries or continents, the
 World Wide Web or our civilization's social networks. In these cases, we
 can focus on only a small portion of the whole network at a given
 moment. The key concept is the one of a roving eye: There is one node in
 focus.  Only this node and nodes within $n$ degrees of separation are
 shown and laid out. Of course we want to shift the focus interactively,
 for example by clicking on a node. The new node in focus defines a new
 set of visible nodes. Dynamic layout then gradually adjusts the
 positions of these nodes to generate a smooth transition from the old
 visualization to the new one.

\begin{figure}[htb]
   \centering
   \includegraphics[height=0.5\textwidth]{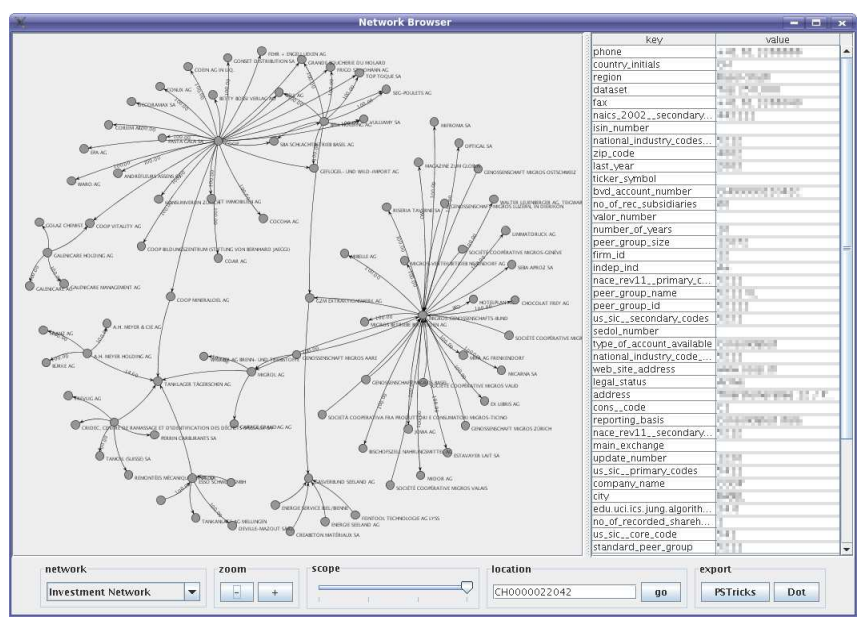}
   \caption{A screen shot of the prototypical network browser. The left
     panel shows a part of the cross investment network of swiss firms,
     surrounding the retailer ``COOP''. The right panel shows firm
     specific details.}
   \label{fig:invest}
 \end{figure}

 Essential for the responsiveness of the interaction are the properties
 of the applied movement model.  The {\em arf} model was successfully
 employed to perform this task.  Large databases containing network data
 can easily be explored. Figure \ref{fig:invest} shows a snapshot. In the
 main window a section of the ownership network of Swiss firms is
 shown\footnote{Based on the ORBIS 2007 Database (Bureau van Dijk.)}. On
 the right more detailed information on the focal node is presented.

 \section{Conclusion}
 \label{sec:conclusion}

 This paper revisited the concept of force directed graph layout and put
 it into a complex systems context. An analysis of the classical spring
 model identified several deficiencies of the resulting layout.  A new
 model was derived to resolve them. A model, especially suited for
 dynamic layout and the visualization of small ($n < 10^3$) networks with
 strongly heterogeneous node degrees. The performance of the new model was
 evaluated and several successful applications in the field of complex
 systems research were discussed.

 Demonstrations and videos can be found at
 \url{http://www.sg.ethz.ch/research/graphlayout}. A
 prototypical open source implementation of the new layouter is also available.
 It can easily be plugged into the open source graph library {\em
   JUNG}\footnote{Website: \url{http://jung.sourceforge.net/}. See also
   \cite{omadadhain05}.}.

 \section*{Acknowledgement}
 I express my gratitude to Frank Schweitzer (Zurich) and Steffano
 Battiston (Zurich) for their valuable comments and suggestions.


\linespread{0.88} \normalsize

\bibliographystyle{abbrv}
\bibliography{Bibliography}

\end{document}